\documentclass[useAMS,twocolumn,usenatbib]{mn2e}
\usepackage{amssymb}
\usepackage{amsmath}
\usepackage{graphicx}
\DeclareGraphicsExtensions{.eps, .ps, .eps.gz, ps.gz}
\usepackage[british]{babel}
\usepackage{placeins}
\usepackage{subfigure}
\usepackage{hyperref}
\usepackage{color}

\begin{document}
%----------------------------- title  -----------------------------

\title{
Measuring the growth of matter fluctuations with third-order galaxy correlations
}
\author[K. Hoffmann, J. Bel, E. Gazta\~naga, M. Crocce, P. Fosalba, F. J. Castander] 
{K. Hoffmann$^{1}$, J. Bel$^{2,3,4}$, E.Gazta\~naga$^{1}$,
M.Crocce$^{1}$, P.Fosalba$^{1}$, F.J.Castander$^{1}$\\
$^{1}$Institut de Ci\`{e}ncies de l'Espai (ICE, IEEC/CSIC), E-08193 Bellaterra (Barcelona), Spain\\
$^{2}$INAF - Osservatorio Astronomico di Brera, Via Brera 28, 20122 Milano, via E. Bianchi 46, 23807 Merate, Italy\\
$^{3}$Aix Marseille Universit\'e, CNRS, Centre de Physique  Th\'eorique, UMR 7332, F-13288, Marseille, France\\
$^{4}$Universit\'e de Toulon, CNRS, CPT, UMR 7332, F-83957 La Garde, France
} 

\date{Received date / Accepted date}

\maketitle

%%%%%%%%%%%%%%%%%%%%%%%%%%%%%%%%%%%%%%%%%%%%%%%%%%%%%%%%%%%%%
%%%%%%%%%%%%%%%%%%%%%%%%%%%%%%%%%%%%%%%%%%%%%%%%%%%%%%%%%%%%%
%%                                                     ABSTRACT                                                           %%
%%%%%%%%%%%%%%%%%%%%%%%%%%%%%%%%%%%%%%%%%%%%%%%%%%%%%%%%%%%%%
%%%%%%%%%%%%%%%%%%%%%%%%%%%%%%%%%%%%%%%%%%%%%%%%%%%%%%%%%%%%%

\begin{abstract}
	Measurements of the linear growth factor $D$ at different redshifts $z$ are key to distinguish among cosmological models.
	One can estimate the derivative $dD(z)/d\ln(1+z)$ from redshift space measurements of the 3D anisotropic galaxy two-point
	correlation $\xi(z)$, but the degeneracy of its transverse (or projected) component with galaxy bias $b$,
	i.e. $\xi_{\perp}(z) \propto\ D^2(z) b^2(z)$, introduces large errors in the growth measurement.
	
	Here we present a comparison between two methods which break this degeneracy by combining second- and third-order
	statistics. One uses the shape of the reduced three-point correlation and the other a combination of third-order one- and two-point cumulants.
	These methods use the fact that, for Gaussian initial conditions and scales larger than $20$ $h^{-1}$Mpc, the reduced third-order matter
	correlations are independent of redshift (and therefore of the growth factor) while the third-order galaxy correlations depend on $b$.
	We use matter and halo catalogs from the MICE-GC simulation to test how well we can recover $b(z)$ and therefore $D(z)$ with these
	methods in 3D real space. We also present a new approach, which enables us to measure $D$ directly from the redshift evolution of
	second- and third-order galaxy correlations without the need of modelling matter correlations.
	
	For haloes with masses lower than $10^{14}$ $h^{-1}$M$_\odot$, we find $10\%$  deviations between the different estimates of
	$D$, which are comparable to current observational errors. At higher  masses we find larger differences that can
	probably be attributed to the breakdown of the bias model and non-Poissonian shot noise.
\end{abstract}
 
\begin{keywords}
large scale structure, clustering, growth, bias, third-order one-, two- and three-point statistics
\end{keywords}

%%%%%%%%%%%%%%%%%%%%%%%%%%%%%%%%%%%%%%%%%%%%%%%%%%%%%%%%%%%%%
%%%%%%%%%%%%%%%%%%%%%%%%%%%%%%%%%%%%%%%%%%%%%%%%%%%%%%%%%%%%%
%%                                     INTRODUCTION (Section I)                                                  %%
%%%%%%%%%%%%%%%%%%%%%%%%%%%%%%%%%%%%%%%%%%%%%%%%%%%%%%%%%%%%%
%%%%%%%%%%%%%%%%%%%%%%%%%%%%%%%%%%%%%%%%%%%%%%%%%%%%%%%%%%%%%
                 
\section{Introduction}

Evidence  that the expansion of the Universe is accelerating  \citep{de1,de2} has
revived  the cosmological constant $\Lambda$, originally introduced by Einstein 
as an unknown fluid which may engine the observed dynamics of the Universe.
Alternative explanations for the accelerated expansion could involve a modification
of the gravitational laws on cosmological scales.
Since  these  modifications of gravity can mimic well the observed accelerated
expansion it is difficult to just rely on the cosmological background 
(i.e. the overall dynamics of the Universe) in order to verify which model is correct.
However, alternative gravitational laws change
the way matter fluctuations grow during the expansion history of our Universe. 
Measuring the growth of matter fluctuations could therefore be a powerful tool to 
distinguish between cosmological models
\citep[see e.g.][]{GL, Lue04, rossetal07, S&P09, C&G09, SP&R12, reidetal12, contrerasetal13, guzzoetal08, delatorreetal13, SB&M}.

On this basis, the goal of several future and ongoing cosmological surveys, such as
BOSS\footnote{https://www.sdss3.org/surveys/boss.php},
DES\footnote{www.darkenergysurvey.org},
MS-DESI\footnote{desi.lbl.gov},
PAU\footnote{www.pausurvey.org},
VIPERS\footnote{http://vipers.inaf.it} or
Euclid\footnote{www.euclid-ec.org},  is to measure the growth 
of matter fluctuations. This can be
achieved by combining several observables, such as weak gravitational lensing, cluster abundance or redshift
space distortions. Higher-order correlations in the galaxy distribution provide additional observables
which also allow for proving the growth equation beyond linear theory from observations \citep[e.g. see][]{bernardeau02}.
Furthermore, higher-order correlations can be used to test the nature
of the initial conditions and improve the signal-to-noise in recovering
cosmological parameters \citep[e.g.][]{Sefusatti06}.

The relative simplicity of the fundamental  predictions about
amplitude and scaling of clustering statistics, must not make us overlook the fundamental difficulty that hampers large scale structure studies. The perfect, continuous (dark matter)
fluid in terms of which we model the large-scale distribution of
matter cannot be directly observed.  Let's imagine that we are able
to locate in the Universe all existing galaxies and that we know with
an infinite precision their masses. Without any knowledge of how
luminous  galaxies trace the underlying continuous distribution of
matter,  even this ultimate galaxy sample would be of limited use.
The problem of unveiling how the density fields of galaxies and mass map into
each other is the so called galaxy biasing. Knowledge of galaxy bias,
and therefore galaxy formation, can greatly improve our cosmological
inferences from observations.

A common approach to model galaxy bias consists in describing the mapping between the fields of mass and galaxy density fluctuations ($\delta_{dm}$ and $\delta_g$ respectively)
by a deterministic local function $F$. This function can be approximated by its Taylor expansion if we smooth the density field on scales that are sufficiently large to ensure that
fluctuations are small,

\begin{equation} 
\delta_g=F[\delta_{dm}] \simeq \sum_{i=0}^{N}\frac{b_i}{i!}\delta_{dm}^i,
\label{eq:biasfunction}
\end{equation}
where $b_i$ are the bias coefficients. It has been shown that,  in this large scale limit,  such a local
transformation preserves the hierarchical properties of matter statistics \citep{FG}. 
There is now convincing evidences about the non-linear character
of the bias function \citep{ga92,mar05, gnbc, mar08, Kovac}.
Since we only want to study correlations up to third order, in this
paper we  shall consider bias coefficients up to second order,
i.e. $b_1$ and $b_2$, which is expected to be sufficient at the leading order \citep{FG}.
However, one of the  goals of this paper is to investigate at which scale and halo mass range
this expectation is fulfilled.

To study the statistical properties of the matter field we need to find the most likely value for the coefficients $b_i$. 
A general approach aims at extracting them from redshift surveys using higher-order statistics.
If the initial perturbations are Gaussian and if the shape of third-order statistics
are correctly described by results of the weakly non-linear perturbation theory, then one  can fix
the amplitude of $b_i$ up to second order in a way which is  independent from the overall amplitude of clustering
(e.g. $\sigma_8$) and depends only on the shape of the linear power spectrum. 
This has been shown by several authors using the
the skewness $S_3$ \citep{gazta94, gaztafrie94},
the bispectrum \citep{Fry94,gaztafrie94,scoc98,feld01,Verde},
the three-point correlation function $Q$, \citep{gnbc, gaztascoc, panszapudi,marin11,mcbride11,marin13},
and the two-point cumulants  $C_{12}$ \citep{bernardeau96,Szapudi98,GFC,bm}.

Recently, \citet{bm} demonstrated that it is possible to use these
higher-order correlations  to constrain bias and 
fundamental properties of the underlying matter field using a
combination of $S_3$ and $C_{12}$, which they call $\tau =3C_{12}- 2S_3$.

The main goal of this paper is to present for the first time a comparison of the bias
derived from this new $\tau$ method with that of $Q$,
using the same simulations and halo samples.
We also show that, with a new approach, the growth of matter fluctuations 
can be measured directly from observations by getting rid of galaxy bias
and without requiring any modelling of the underlying matter distribution.
Despite the fact that in the present analysis we only consider
real-space observables (not affected by redshift-space distortions)
we argue that, as long as reduced third-order statistics are only
weakly affected by redshift-space distortions (for a broad range of
masses, see Fig. \ref{fig:q3pz}), the proposed method appears to be
applicable on redshift galaxy surveys.

This analysis is based on the new MICE-GC simulation and extends its validation
presented recently by \citet{mice1, mice2,  mice3}.

In Section \ref{sec:micegc} we present the simulation on which our work relies.
Our estimators for both, the bias and the growth of matter fluctuations, are introduced in Section \ref{sec:methods}.
We present our results in Section \ref{sec:results} and a summary of the work
can be found in Section \ref{sec:disc} together with our conclusions.

%%%%%%%%%%%%%%%%%%%%%%%%%%%%%%%%%%%%%%%%%%%%%%%%%%%%%%%%%%%%%
%%%%%%%%%%%%%%%%%%%%%%%%%%%%%%%%%%%%%%%%%%%%%%%%%%%%%%%%%%%%%
%%                                             SIMULATION (Section II)                                              %%
%%%%%%%%%%%%%%%%%%%%%%%%%%%%%%%%%%%%%%%%%%%%%%%%%%%%%%%%%%%%%
%%%%%%%%%%%%%%%%%%%%%%%%%%%%%%%%%%%%%%%%%%%%%%%%%%%%%%%%%%%%%

\section{Simulation and halo samples}\label{sec:micegc}

Our analysis is based on the Grand Challenge run of the  Marenostrum
Institut de Ci\`encies de l'Espai (MICE) simulation suite to which we refer to
as MICE-GC in the following.
Starting from small initial density fluctuations at redshift $z=100$ the formation of large scale cosmic structure was computed with $4096^3$ gravitationally
interacting collisionless particles in a $3072$ $h^{-1}$Mpc box using the GADGET - 2 code \citep{springel05} with a softening length of $50$ $h^{-1}$kpc. The initial conditions
were generated using the Zel'dovich approximation and a CAMB power spectrum with the power law index of $n_s = 0.95$, which was normalised to be
$\sigma_8 = 0.8$ at $z=0$.  The cosmic expansion is described by the $\Lambda$CDM model for a flat universe with a mass density of
$\Omega_m$ = $\Omega_{dm} + \Omega_b = 0.25$. The density of the
baryonic mass is set to $\Omega_b = 0.044$ and $\Omega_{dm}$ is the dark matter density.
The dimensionless Hubble parameter is set to $h = 0.7$. 
More details and validation test on this simulation can be found in \citet{mice1}.

Dark matter haloes were identified as Friends-of-Friends groups \citep{davis85} with a linking length of $0.2$ in units of the mean particle separation.
These halo catalogs and the corresponding validation checks are
presented in \citet{mice2}.

To study the galaxy bias and estimate the growth as a function of halo
mass  we divide the haloes into the four redshift independent mass
samples M0, M1, M2 and M3, shown in Table \ref{table:halo_masses}.
They span a mass range from Milky Way like haloes (M0) up to massive galaxy clusters (M3).
In the same table we show the total number and comoving number density of haloes at redshift $z=0.5$, 
a characteristic  redshift for current galaxy surveys, such as BOSS LRG.

\begin{table}
\centering
  \caption{Halo mass samples. $N_p$ is the number of particles per halo, $N_{halo}$ is the number of haloes per sample
  in the comoving output at $z=0.5$.  $n_{halo}$ is the comoving number density of haloes.
  $N_{halo}$ and $n_{halo}$ are compared to the corresponding values in the light cone in Fig. \ref{fig:nofz}.}
  \label{table:halo_masses}
  \begin{tabular}{c  c c c c}
 \hline
   & mass range  &  $N_p$ & $N_{halo}$ & $n_{halo}$\\
 &  $10^{12} M_{\odot}/h$  & & & $(10$ Mpc$/h)^{-3}$\\
     \hline 
 M0	&	$0.58-2.32$	&	$20-80$		&	$122300728$	&	$4.22$ \\ 
 M1	&	$2.32-9.26$	&	$80-316$		&	$31765907$	&	$1.10$ \\ 
 M2	&	$9.26-100$	&	$316-3416$	&	$8505326$	&	$0.29$ \\ 
 M3	&	$\ge100$		&	$\ge3416$	&	$280837$		&	$0.01$ \\ 
      \hline
   \end{tabular}
\end{table}

\begin{figure}
   \centering
   \includegraphics[width=3.4in, angle = 270]{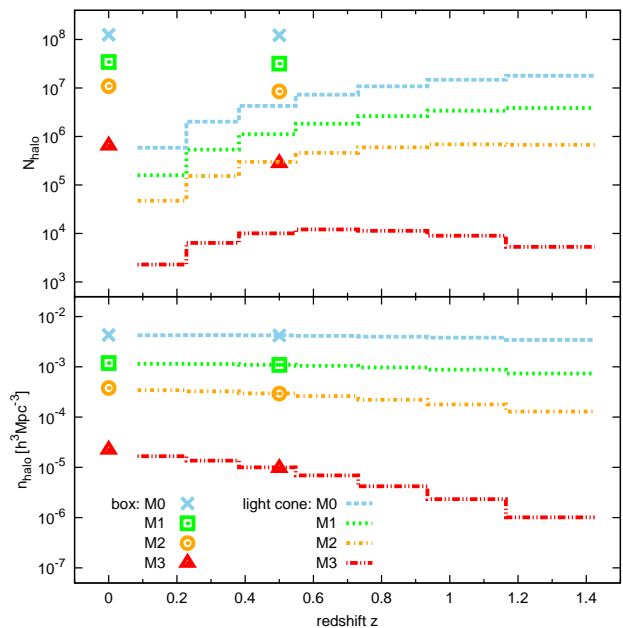}
   \caption{{\it Top:} Number of haloes in the four mass samples M0-M3 as a
     function of redshift in the two comoving outputs at z=0.0 and z=0.5 (symbols) and
     the seven redshift bins in the light cone (lines).{\it Bottom:} number density of the
     same halo mass samples as in the top panel.}
   \label{fig:nofz}
\end{figure}

We are analysing two types of simulation outputs. For a detailed study
of the dark matter growth we use the full comoving output at redshift
$z = 0.0, 0.5, 1.0$  and $1.5$. 
For studying the bias estimators with minimal shot noise and sampling
variance we use haloes identified in the comoving outputs at  redshift
$z = 0.0$ and $0.5$. 
The investigation of the redshift evolution of the bias and growth
estimators is based  on seven redshift bins of the light cone output 
with equal width of $400$ $h^{-1}$Mpc in comoving space over one octant of the sky.
Fig. \ref{fig:nofz} shows the number and number density of haloes in the
four mass samples for the comoving output and the light cone with respect to the redshift.

%%%%%%%%%%%%%%%%%%%%%%%%%%%%%%%%%%%%%%%%%%%%%%%%%%%%%%%%%%%%%
%%%%%%%%%%%%%%%%%%%%%%%%%%%%%%%%%%%%%%%%%%%%%%%%%%%%%%%%%%%%%
%%                                       GROWTH & BIAS (Section III)                                             %%
%%%%%%%%%%%%%%%%%%%%%%%%%%%%%%%%%%%%%%%%%%%%%%%%%%%%%%%%%%%%%
%%%%%%%%%%%%%%%%%%%%%%%%%%%%%%%%%%%%%%%%%%%%%%%%%%%%%%%%%%%%%

\section{Growth and Bias Estimators}\label{sec:methods}

%%%%%%%%%%%%%%%%%%%%%%%%%%%%%%%%%%%%%
%                              Sub-Section I                           %
%%%%%%%%%%%%%%%%%%%%%%%%%%%%%%%%%%%%%

\subsection{The growth factor}\label{sec:growth_factor}

The large scale structure in the distribution of galaxies, observed today
in cosmological surveys,  is believed to originate from some small initially
gaussian matter density fluctuations that grew with time due to
gravitational instabilities.
Since the way the Universe is expanding has an impact of the
  growth of structures, one can use measurements of the growth to put
  constraints on various cosmological models.

We adopt the common definition for density fluctuations, given by 
$\delta(\boldsymbol{r}) = \rho(\boldsymbol{r}) /\overline \rho-1$,
where $\rho(\boldsymbol{r})$ is the density at position
$\boldsymbol{r}$ smoothed  (with a spherical top-hat window) over the radius 
$R$, while $\overline \rho$ is the mean density of the Universe. 
In the linear regime (large smoothing scales) density fluctuations of matter $\delta_m(
\boldsymbol{r},z)$ evolve with the redshift $z$ 
in a self similar way, thus
\begin{equation}
\delta_m (\boldsymbol{r},z) = D(z) ~\delta_m (\boldsymbol{r},z_0).
\label{eq:def_growth}
\end{equation}
The reference redshift $z_0$ is usually arbitrarily
chosen to be today, i.e. $z_0=0$.
In the $\Lambda$CDM model the growth factor $D(z)$ depends on
cosmological parameters via the Hubble expansion rate

\begin{equation}
H(z)=H(0)\sqrt{ \Omega_m (1+z)^{3}+(1-\Omega_m -\Omega_\Lambda)(1+z)^{2}  +\Omega_\Lambda },
\label{hrate}
\end{equation}
where $\Omega_{m}$ and $\Omega_{\Lambda}$
are the densities of matter and dark energy respectively, and the
growth factor is then given by:
\begin{equation}
D(z) \propto  H(z)\int_z^\infty \frac{1+z'}{ H^3(z') }dz'.
\label{eq:growth_cosmology}
\end{equation}
However, in general $D$ is also sensitive
to modifications of the gravity action on cosmological scales (e.g. see
Gazta\~naga \& Lobo 2001 and references therein).
Measurements of the growth factor as a  function of redshift can therefore 
be used to constrain cosmological models and understand the nature of
cosmic expansion.

In practice, instead of computing the integral equation (\ref{eq:growth_cosmology}), one can also approximate the growth factor
with an analytic expression, which is accurate at any redshift for a flat $\Lambda$CDM model.
This approximations can be obtained in two steps. First, on deriving
equation (\ref{eq:growth_cosmology}) one can express the growth 
factor in terms of the growth rate

\begin{equation}
 f(z) \equiv\frac{d \ln D}{d \ln a},
\label{eq:f(z)}
\end{equation}
where $a=1/(1+z)$. It follows that  

\begin{equation}
D(z) \propto \frac{(1+z)^2}{H^2(z)}\left\{ f(z) + 1 +
  \frac{\Omega_m(z)}{2} - \Omega_\Lambda(z) \right\}^{-1},
\label{growing}
\end{equation}
which is an exact solution for any $\Lambda$CDM (i.e. can be characterised by
a curved space) cosmological model. 
Second, for a spatially flat universe, we can use the growth index $\alpha$ defined as

\begin{equation}
f(z) \equiv \left[ \Omega_m(z)\right]^{\alpha(z)}. 
\label{growing2}
\end{equation}
\citet{WS98}  found that it can be approximated by
$$\alpha(z) \simeq \frac{6}{11} + \frac{30}{2662}[1-
\Omega_m(z) ] 
$$
 which provides a relative precision of $0.2$\% on the growth factor.
Recently \citet{SB&M} found an even more precise expression
$$
\alpha(z) \simeq \frac{6}{11} - \frac{15}{2057}\ln[\Omega_m(z)] +  \frac{205}{540 421}\ln^2[\Omega_m(z)],
$$ 
which increases the accuracy of equation (\ref{growing}) to better then $0.01$\% at all redshift ($0\leq z \leq 100$).

Measuring the growth factor using equation (\ref{eq:def_growth})
requires knowledge of the matter density fluctuations $\delta_m$ at different redshifts, while in practice only galaxies can be observed
as biased tracers of the matter field. In the following sections we describe how we quantify and measure this galaxy bias.

%%%%%%%%%%%%%%%%%%%%%%%%%%%%%%%%%%%%%
%                              Sub-Section II                          %
%%%%%%%%%%%%%%%%%%%%%%%%%%%%%%%%%%%%%

\subsection{The local bias model}

Our bias estimations are based on the local bias model (Fry \& Gaztanaga 1993), which assumes that the 
galaxy (number density) fluctuation $\delta_g$ is 
a function of the matter density fluctuation $\delta_{dm}$ at the same location: $\delta_g=F[\delta_{dm}]$, while both fluctuations
are smoothed at the same scale $R$. For sufficiently large smoothing scales the density fluctuations become small and we can expand this
function, i.e. as in equation (\ref{eq:biasfunction}). For third-order statistics it is enough to stop the expansion at quadratic order (e.g. see Fry \& Gaztanaga 1993)
\begin{equation}
\delta_g = b_1  \left\lbrace \delta_{dm} + \frac{c_2}{2}(\delta_{dm}^2 - \langle \delta_{dm}^2  \rangle) \right\rbrace,
\label{eq:def_biasmodel}
\end{equation}
where $b_1$ and $c_2$ are, respectively, the linear and quadratic bias parameters which we are measuring. The term $\langle \delta_{dm}^2 \rangle$ ensures that $ \langle \delta_g  \rangle = 0$,
where $\langle \dots \rangle$ denotes the average over all spatial positions.
Besides small density fluctuations, such a model for the bias assumes that neither the environment nor the velocity field has an impact on galaxy formation.

Recent studies have shown, that the local assumption might not be 
accurate for small smoothing scales when $b_1$ is large \citep{baldauf12,chan12}.

Using the information contained in the large scale distribution of galaxies at different scales we measure bias and growth with second- and third-order statistics, as described in the following sections.

%%%%%%%%%%%%%%%%%%%%%%%%%%%%%%%%%%%%%
%                              Sub-Section III                          %
%%%%%%%%%%%%%%%%%%%%%%%%%%%%%%%%%%%%%

\subsection{Growth factor $D$ from two-point correlation $\xi$}\label{sec:Dfrom2pc}

\begin{figure}
\centerline{\includegraphics[width=105mm,angle=270]{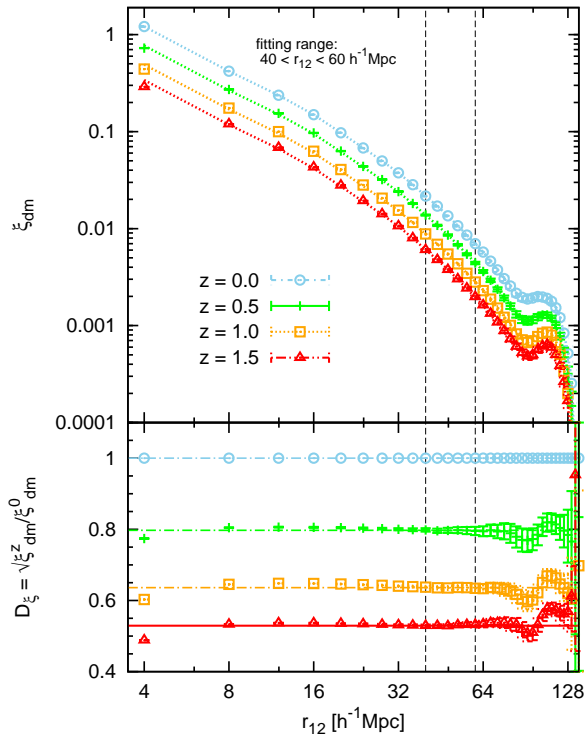}}
\caption{  {\it Top:} two-point correlation $\xi$ of the MICE-GC dark matter field
	  measured in the comoving outputs at redshift $z=0.0$, $0.5$, $1.0$ and $1.5$ (blue
	  circles, green crosses, orange squares and red triangles respectively) as a function
	  of scale $r_{12}$. Dotted Lines show a fit of the amplitude of $\xi$ at $z=0$ to those
	  from other redshifts between $40-60$ $h^{-1}$Mpc, via equation (\ref{ximgrow}).
	  {\it Bottom:} growth factor $D=\sqrt{\xi(r_{12},z)/\xi(r_{12},0)} $ obtained from the
	  ratio of the above correlations together with the fits displayed as dotted lines  with
	  the same colour code as the upper panel.}
\label{fig:D_from_2pc}
\end{figure}
\begin{figure}
\centerline{\includegraphics[width=75mm,angle=270]{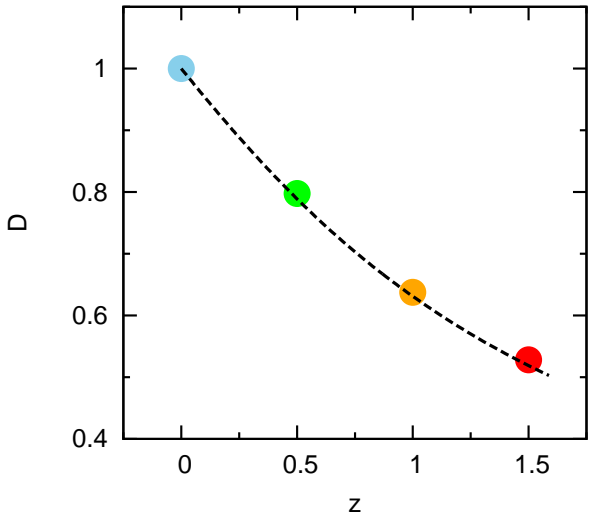}}
\caption{Comparison between the linear growth of matter $D$ as a function of redshift $z$
measured in the MICE-GC comoving outputs (symbols) and the corresponding theoretical
predictions from equation (\ref{eq:growth_cosmology}) (dashed line). The MICE-GC
measurements are the best fit values obtained considering the scale range $40$-$60h^{-1}$Mpc,
shown as lines with the same colour coding in Fig. \ref{fig:D_from_2pc}.}
\label{fig:D_2pcvsPT}
\end{figure}

The spatial two-point correlation of density fluctuations can be defined as the mean product of density fluctuations $\delta_i$ at the positions $\boldsymbol{r_i}$ that are separated by the distance
$r_{12} \equiv \vert \boldsymbol{r_1}-\boldsymbol{r_2} \vert$,
\begin{equation}
\xi(r_{12}) \equiv \langle  \delta(\boldsymbol{r_1})
\delta(\boldsymbol{r_2}) \rangle =
\langle \delta_1 \delta_2 \rangle(r_{12}).
\label{eq:def_2pc}
\end{equation}
Note that the two-point correlation function depends only on the separation 
and thus is not sensitive to the shape of over-densities.
This is in contrast with higher-order correlations: with three points we will
also be able to measure deviations away from the spherically symmetric profile
\citep{smith06}.

From equations (\ref{eq:def_growth}) and (\ref{eq:def_2pc}) 
one can derive that the growth factor is related to the  two-point correlation
of matter as 

\begin{equation} 
\xi_{dm}(r_{12},z) = D(z)^2 \xi_{dm}(r_{12}, z_0).
\label{ximgrow}
\end{equation}

Our measurements of the matter correlation function in the MICE-GC
simulation,  presented in Fig. \ref{fig:D_from_2pc}, indeed show a linear relation between the
matter two-point correlations at different redshifts $z$ with respect to $z_0=0$
on a wide range of scales. Note how at scales around $r_{12}\sim 100$
$h^{-1}$Mpc the BAO peak induces some oscillations around the linear model,
but the model works well for intermediate scales of $r_{12}\sim 40-60$ $h^{-1}$Mpc.

We measured the two-point correlation by dividing the simulation volume into cubical grid cells and
assigning density fluctuations to each of these cells. We then calculate the mean product of density
fluctuations in grid cells that are separated by $r_{12} \pm dr$ according to definition (\ref{eq:def_2pc}).
The measurements shown in Fig. \ref{fig:D_from_2pc} are based on $4$ $h^{-1}$Mpc grid cells. Errors are derived by
Jackknife resampling as described in Section \ref{sec:err}.

As shown in  Fig. \ref{fig:D_2pcvsPT}, there is a good agreement between the growth factor measurements
from the two-point correlation (symbols) and the theoretical prediction from equation (\ref{eq:growth_cosmology})
(dashed line) for the cosmology of the MICE-GC simulation.

Deviations between predictions and measurements are at the sub-percent level and result from to non-linearities
\footnote{The increase of the deviations with redshift results from non-linearities in $\xi_{dm}$ at $z=0.0$. Since we
use the latter as normalisation, deviations between predictions and measurements transfer to the higher redshifts. If we use $\xi$
at high $z$ as normalisation, as we do it later in the light cone, this effect goes into the opposite direction.}.
This result demonstrates that, in principle, we can obtain constrains on cosmological models by just measuring the
two-point correlation function of matter.

However such constraints are difficult to realise as we
have to infer the correlation of the unobservable full matter field from the correlation of the
observed galaxy distribution.
A simple relation between the two-point correlation functions of matter and
galaxies can be obtained by inserting the model for galaxy bias, given
by equation (\ref{eq:def_biasmodel}), into the definition of the
two-point correlation (equation \ref{eq:def_2pc}). 
At leading order

\begin{equation}
\xi_g(r_{12}, z)  \simeq ~b_1^2~\xi_{dm}(r_{12},z) + {\cal O}[\xi_{dm}^2].
\end{equation}
This relation
only holds for sufficiently large separations $r_{12}$, where we
can neglect terms of order  $\xi_{dm}^2$, and small
density fluctuations $\delta_{dm}$ in equation
(\ref{eq:def_biasmodel}).  To estimate  the linear bias from $\xi$ we define
\begin{equation}
b_{\xi}(z) \equiv\sqrt{ \frac{\xi_g(r_{12}, z)}{\xi_{dm}(r_{12},z)} }
\simeq b_1
\label{bxi}
\end{equation}
which is expected to be independent of separation in the large scale limit. 
In the following sections we will estimate $b_1$ from two other estimators ($b_Q$ and $b_\tau$) based on third-order statistics and we will probe how they compare to each other.

\begin{figure*}
   \centering
   \includegraphics[width=110mm, angle = 270]{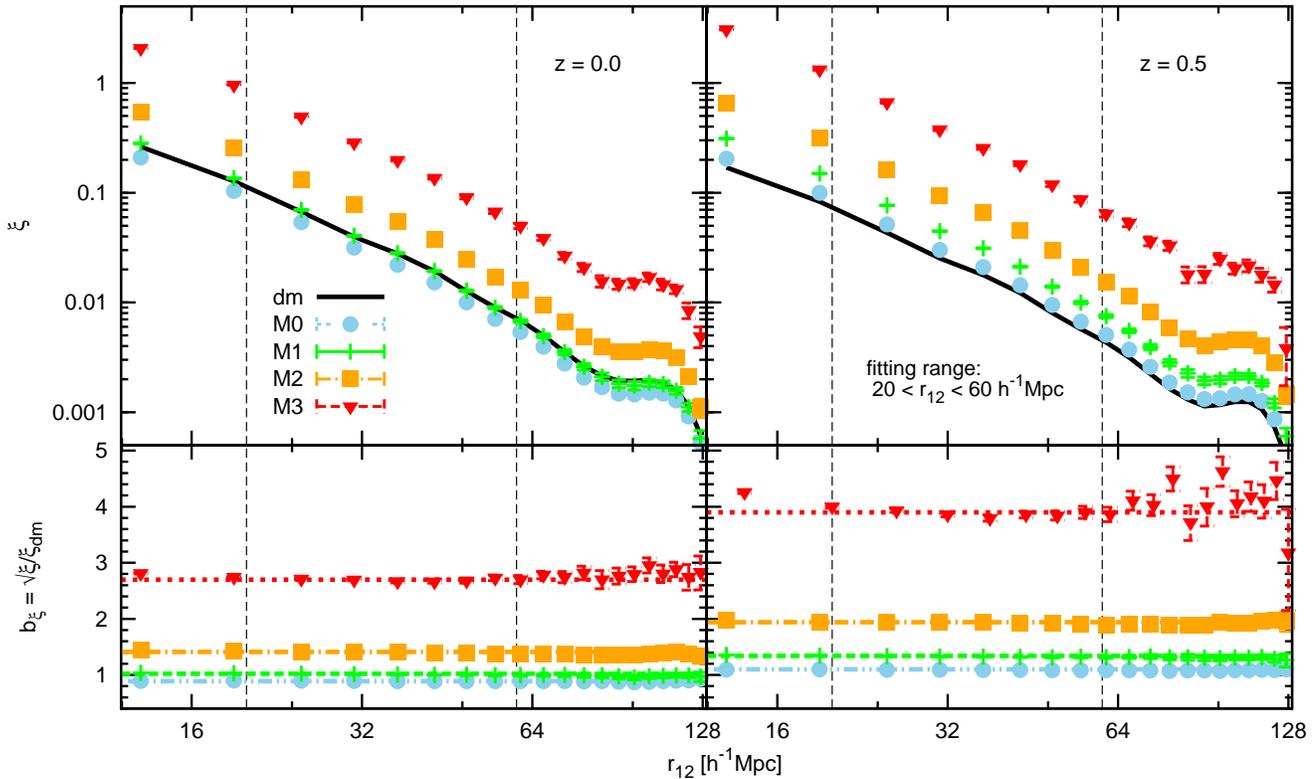}
   \caption{\emph{Top}: two-point correlation $\xi$ of the MICE-GC
     dark matter field (continuous lines) and the four halo mass samples M0-M3
   (blue circles, green crosses, orange squares and red triangles respectively) in the comoving outputs
   at redshift $z=0.0$ (left) and $z=0.5$ (right) as a function of scale $r_{12}$.
   \emph{Bottom}: linear bias parameter $b_{\xi}$ derived from the two-point correlations via equation (\ref{bxi}).
   Dotted lines are $\chi^2$-fits between $20-60$ $h^{-1}$Mpc.
   The minimum $\chi^2$ values per degree of freedom are
   $1.05, 2.02, 0.37, 0.70$ for M0, M1, M2, M3 respectively at $z=0.0$ and
   $0.42, 0.78, 0.12, 0.82$ for M0, M1, M2, M3 respectively at $z=0.5$.}
   \label{fig:b_from_2pc}
\end{figure*}

The correlation functions of the halo samples M0 - M3, calculated with $8$ $h^{-1}$Mpc grid cells, are shown in the
top panel of Fig. \ref{fig:b_from_2pc} together with the corresponding measurements for the dark matter field. The ratios of the matter and
halo correlations, shown in the bottom panel, confirm that both quantities can be related by the scale independent bias factor $b_{\xi}$
between $20\lesssim r_{12}\lesssim60h^{-1}$Mpc. We expect that $b_{\xi}(r_{12}) \simeq b_1$ at the mass and scale range of our analysis \citep{mice2}.
To estimate $b_{\xi}$ we perform a $\chi^2$-fit to the ratio of the halo and matter two-point correlation in the aforementioned scale range
(see Subsection \ref{sec:err} for details). We find the $\chi^2_{min}$ values to vary between $2.0$ and $0.1$. Values are smaller
at $z=0.5$ compared to $z=0.0$. Restricting the fitting range to larger scales ($30\lesssim r_{12}\lesssim60h^{-1}$Mpc) also reduces
the $\chi^2_{min}$. Both findings are expected since non-linearities enter equation (\ref{bxi}) at small scales and low redshift. However,
restricting the fit to larger scales, as mentioned before, causes a maximum change in bias values is $1.5$ percent. We therefore
consider our $b_{\xi}$ measurements as relatively robust, compared to the bias measurements from higher order statistics.
The fitted bias factors (bottom panel of Fig. \ref{fig:b_from_2pc}) reveal the well
known increase of bias with the mass and redshift of the halo samples.

The results described above allow us to estimate the growth factor of matter
fluctuations from equation (\ref{ximgrow}) in terms of galaxy (or halo) correlation functions as:
\begin{equation}
D (z) \simeq \hat{b}(z)^{-1} D_g(z),
\label{eq:D_from_2pc}
\end{equation}
where the growth factor is normalised to unity at an arbitrary redshift $z_0$ (i.e. $D(z_0)\equiv 1$). The bias ratio $\hat{b}(z)$  is defined as

\begin{equation}
\hat{b}(z) \equiv b(z)/b(z_0)
\label{un}
\end{equation}
and the galaxy (or halo) growth factor $D_g(z)$ is:

\begin{equation}
D_g(z) \equiv \sqrt{\frac{\xi_g(z)}{\xi_g(z_0)}}.
\label{deux}
\end{equation}
Both definitions (\ref{un}) and (\ref{deux}) refer to large scales, i.e $r_{12}$ between $20$-$60$ $h^{-1}$Mpc, while we find changes in the results at the percent level when we vary the fitting range.
The bias at the two different redshifts $z$ and $z_0$ does not need to refer to the same galaxy (or halo) populations. In Section \ref{d_comparison} we
demonstrate that taking different halo masses across the explored redshift range does not lead to unexpected growth measurements.

Equation (\ref{eq:D_from_2pc}) shows that the matter growth factor,
measured from the  galaxy (or halo)
two-point correlation functions at different redshifts is fully degenerate
with the ratio of the linear bias parameters.  
We therefore need an independent measurement of the bias ratio to
break this degeneracy. 

Note that the absolute values of the bias parameters, $b(z)$ and $b(z_0)$, do not need to be measured separately
for measuring the differential growth factor between two redshift bins, as it is commonly done.
Instead of the absolute bias values, we only need to measure their ratio $\hat{b}$, which can be
obtained directly from third-order galaxy correlations without assumptions on the clustering of dark matter,
as we will explain in Subsection \ref{sec:3pc}.

By measuring the differential growth factor between two nearby redshift bins $z_2$ and $z_1$ one can also estimate
the (velocity) growth rate $f(z)$ defined in equation (\ref{eq:f(z)}) at the mean redshift $\bar{z} \equiv \frac{z_1+z_2}{2}$.
Since the growth rate is defined as logarithmic derivative of the growth factor, it follows that 

\begin{equation}
\begin{array}{rl}
\displaystyle  f(\bar z) & \displaystyle \simeq -\frac{\ln[ D(z_2)/D(z_1)]}{\ln[ (1+z_2)/(1+z_1)]}\\ \\
\displaystyle              & \displaystyle \simeq -\frac{\ln[ b(z_1)/b(z_2) D_g(z_2)/D_g(z_1)]}{\ln[ (1+z_2)/(1+z_1)]}
\end{array}
\label{growratedis}
\end{equation}
Our new approach of measuring the bias ratio $\hat{b}$ with
third-order galaxy correlations will enable us to measure the growth factor and the growth rate of the full matter
distribution directly  from the distribution of galaxies (or haloes) without assumptions on the clustering of dark matter, providing a new model independent constrain
on cosmological parameters.
The same approach represents an additional tool to measure $f(z)$, which is independent of redshift space distortions method \citep{ka87}.
Note that we do not need to select the same type of objects (with respect to the halo mass) at the different redshifts. This  feature allows for
maximisation of the galaxy number density at each redshift.

%%%%%%%%%%%%%%%%%%%%%%%%%%%%%%%%%%%%%
%                              Sub-Section IV                         %
%%%%%%%%%%%%%%%%%%%%%%%%%%%%%%%%%%%%%

\subsection{Bias $b_Q$ from the three-point correlation $Q$}\label{sec:3pc}

In  analogy to the two-point correlation,  we can define the three-point correlation as

\begin{equation}
\zeta (r_{12}, r_{13}, r_{23}) \equiv \langle  \delta(\boldsymbol{r_1})
\delta(\boldsymbol{r_2}) \delta(\boldsymbol{r_3}) \rangle,
\label{eq:def_3pc}
\end{equation}
where the vectors $\bf{r}_{12}, \bf{r}_{13}, \bf{r}_{23}$ form triangles of different shapes and sizes.
In contrast to the two-point correlation function $\zeta$  is
sensitive to the shape  of the matter density fluctuations.
To access this additional information, we 
fix the length of the two triangle legs $r_{12}$ and $r_{13}$
while varying the angle between them, $\alpha = acos(\hat{\bf r}_{12} \cdot \hat{\bf r}_{13} )$. In the following
we will therefore change the variables for characterising triangles from $(r_{12}, r_{13}, r_{23})$ to $(r_{12}, r_{13}, \alpha)$.
Throughout the analysis we use triangles with $r_{13}/r_{12} = 2$ configurations, which restricts the minimum scale entering the
measurements to the size of the smaller triangle leg $r_{12}$. Choosing configuration, such as $r_{13}/r_{12} = 1$ would introduce
non-linear scales when triangles are collapsed ($\alpha = 0$).

For detecting the triples $\delta(\boldsymbol{r_1})\delta(\boldsymbol{r_2}) \delta(\boldsymbol{r_3})$ we employ the algorithm described by \cite{bargaz},
using the same kind of mesh as for calculating the two-point correlation with $4$ and $8$ $h^{-1}$Mpc grid cells.
From the three-point correlation we then construct the reduced
three-point correlation,  introduced by
\cite{grope77} as
\begin{equation}
	Q \equiv \\
	\frac{\zeta(r_{12}, r_{13}, \alpha)}{ \xi_{12} \xi_{13}  + \xi_{12} \xi_{23}  + \xi_{13} \xi_{23} },
	\label{eq:def_Q3}
\end{equation}
where $\xi_{ij} \equiv \xi(r_{ij})$.

Perturbation theory shows that, to leading order in the dark matter field, $Q$ (hereafter referred to as Qdm) is  independent of the growth factor. This is because 
{\bf  $\zeta \propto  \langle \delta_L \delta_L \delta_L^2 \rangle \propto D^4$}, so $D$ drops in the $Q$ ratio above  \citep{bernardeau94, K&B99},
but for galaxies $Q$ depends on the bias parameters.
These properties enable us to measure $b_1$ and $c_2$ and  break the growth-bias degeneracy
in equation (\ref{eq:D_from_2pc}) \citep{friga1994,Fry94, bernardeau02}. 

\begin{figure}
   \centering 
   \includegraphics[width=100mm, angle = 270]{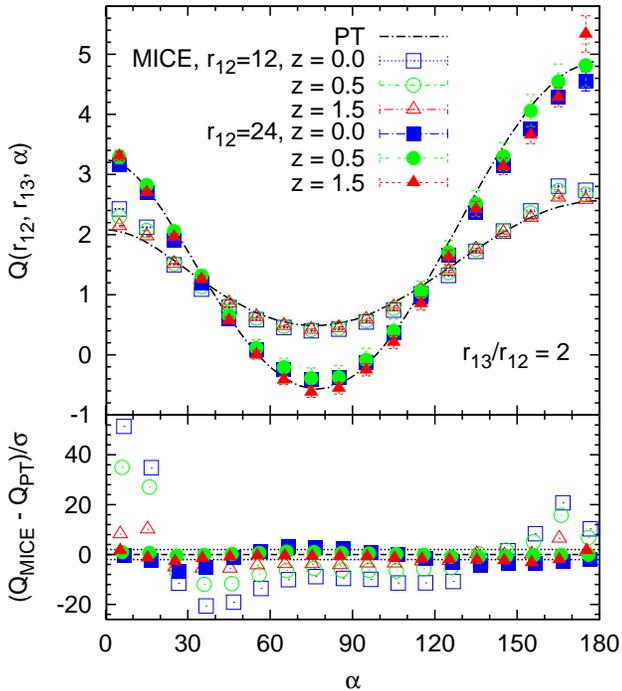}
   \caption{\emph{Top}: reduced three-point correlation $Q$ measured from the MICE-GC
   			dark matter field in the comoving outputs at
                        redshift $z=0.0,0.5,1.5$
			 (blue squares, green circles, red triangles respectively)
			for different triangle opening angles $\alpha$ using $r_{12}=r_{13}/2=12$ $h^{-1}$Mpc (open
                        symbols) and   $r_{12}=r_{13}/2=24$ $h^{-1}$Mpc  (filled symbols)
			compared with predictions from second-order perturbation theory (PT)
   			using a linear power spectrum. \emph{Bottom:}
                        Deviations between $Q$ from PT and
                        measurements divided by the $1 \sigma$ errors
                        of the measurements (dashed lines correspond to
                        $\pm 2\sigma$ discrepancies).}
   \label{fig:3pc1}
\end{figure}

We test the assumption that $Q_{dm}$ is independent of
the growth factor by comparing measurements at different redshifts and scales in the MICE-GC simulation
with theoretical predictions derived from second-order perturbative expansion of $\xi$ and $\zeta$ \citep{bernardeau02,bargaz}.
The predictions are based on the MICE-GC CAMB linear power spectrum.
Fig. \ref{fig:3pc1} shows $Q_{dm}$ at $z=0.0$, $0.5$ and $1.5$ for triangles with $r_{12} = 12$ $h^{-1}$ Mpc and $r_{13} = 24$ $h^{-1}$Mpc.
The measurements are based on a density mesh with $4$ $h^{-1}$Mpc grid cells, which is the highest available resolution (see Table \ref{triangle_configs} for details).
As for the two-point correlation we derive errors for $Q$ by Jackknife resampling (see Section \ref{sec:err}).
The values of $Q$ show the characteristic u-shape predicted by perturbation theory, which results from the
%anisotropic matter distribution
anisotropy of the shape of matter fluctuations. The amplitude of $Q$ increases with triangle size because of the steeper slope in the two-point linear correlations at larger scales.
Also $Q$ depends only weakly on redshift while deviations between predictions and measurements become more significant at
low redshift and small scales (see bottom panel of Fig. \ref{fig:3pc1}). The same effect has been reported by \citet{mice1}, who also find that the
deviations decrease, when predictions are drawn from the measured instead of the CAMB power spectrum. Furthermore, these authors demonstrated
that additional contributions to these deviations can result from the limited mass resolution of the simulation, especially at small scales and high redshift.

\subsubsection{Non-linear bias}

\begin{figure*}
  \centering
   \includegraphics[width=88mm, angle = 270]{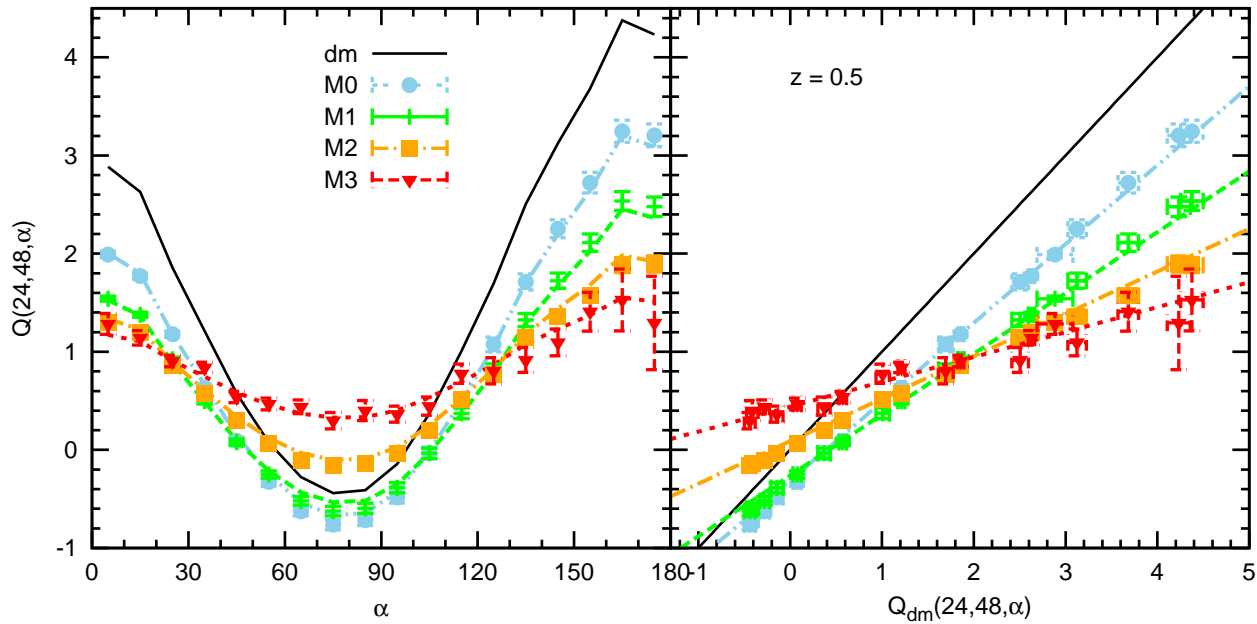}
   \caption{\emph{Left}: reduced three-point correlation $Q$ measured from the MICE-GC dark matter field and the four halo mass samples M0-M3
   (black line, blue circles, green crosses, orange squares and red triangles respectively) in the comoving output at redshift $z = 0.5$ 
   for different triangle opening angles $\alpha$ using triangles with $r_{12} = 24$ $h^{-1}$Mpc and $r_{13} = 48$ $h^{-1}$Mpc.
   \emph{Right} : $Q_{dm}$ versus $Q_g$ at the corresponding opening angle with the same colour coding as in the left panel.
    Dashed lines are $\chi^2$-fits to equation (\ref{eq:b1c2_q3auto}). The minimum $\chi^2$ per degree of freedom is
    $6.0$, $3.9$, $2.0$, $0.7$ for M0, M1, M2 and M3 respectively. Results for redshift $z=0.0$ are shown in the Appendix.}
   \label{fig:3pc2}
\end{figure*}

A simple relation between the bias in the local model and $Q$ can be
derived in the limit of small density fluctuations and large triangles by using
equation (\ref{eq:def_biasmodel}) with the definitions (\ref{eq:def_2pc}),
(\ref{eq:def_3pc}) and (\ref{eq:def_Q3}), and keeping second-order
terms in the perturbative expansion \citep{friga1994}:
\begin{equation}
	Q_g(\alpha) \simeq \frac{1}{b_Q} [Q_{dm}(\alpha)+c_Q].
	\label{eq:b1c2_q3auto}
\end{equation}
Instead of using $Q_{dm}$, we could also use the corresponding predictions, shown in Fig. \ref{fig:3pc2}.
However, this would introduce uncertainties in the bias measurement, due to the mismatch between
measurements and predictions.
We interpret the parameters $b_Q$ and $c_Q$ as the first- and second-order bias
parameters $b_1$ and $c_2$ respectively, while we expect this
interpretation to be valid only in the linear regime at scales larger
than roughly $20$ $h^{-1}$Mpc. We use the notation $b_Q$ instead of
$b_1$ to refer to the fact that we are estimating $b_1$ with $Q$.

To measure the bias we computed $Q_g$ for the four mass samples M0 - M3 at redshift $z=0.0$ and $z=0.5$
using triangles of various scales with  $r_{13}=2 r_{12}$ configurations. The triangle legs consist now of $3$ and $6$
$\pm0.5$ grid cells (see Table \ref{triangle_configs} for details). We vary the size of the triangles by changing
the size of the grid cells. This reduces computation time, since the number of grid cells in the simulation volume required for the measurement is minimised.
Our bias measurements from $Q$ can vary by less than $5\%$, when we increase the number of cells per leg instead of increasing the cell size
to measure $Q$ at larger scales. We show and discuss this effect in Appendix \ref{sec:app}.

Our results for   $r_{12}=r_{13}/2=24$ $h^{-1}$Mpc triangles, shown in the left panel of Fig. \ref{fig:3pc2} reveal a
flattening of $Q$ for high mass samples, as expected from equation (\ref{eq:b1c2_q3auto}) since
$b_1$ increases with halo mass. In the right panel of the same figure we demonstrate that the linear relation between $Q_g$ and $Q_{dm}$, given by
equation (\ref{eq:b1c2_q3auto}) is in reasonable agreement with the measurements.
We perform $\chi^2$-fits of the dark matter results to those of the four halo samples via equation (\ref{eq:b1c2_q3auto})
as described in Subsection \ref{sec:err} and obtain the bias parameters $b_Q$ and $c_Q$.  
These fits, shown as colored lines in Fig. \ref{fig:3pc2}, have the strongest deviations from the measurements
at the smallest and highest values of $Q$, which might result from measurements at small angles
dominating $\chi^2$ as those have the smallest errors. The corresponding minimum values of $\chi^2$ per degree of
freedom (given in the caption of Fig. \ref{fig:3pc2}) decrease for higher mass samples as the errors of $Q$ increase.
In general we find a decrease with mass, scale and redshift. Note that these results are affected by the covariance
matrix in the fit which we only know roughly from the Jackknife sampling (see Section \ref{sec:err} and Appendix \ref{sec:app}).

\begin{figure}
   \centering
   \includegraphics[width=95mm, angle = 270]{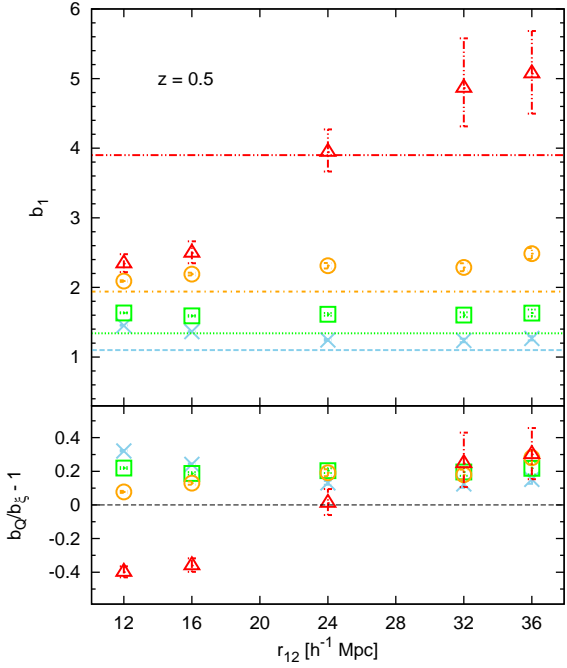}
   \caption{\emph{Top}: linear bias parameter $b_1$ derived from the three-point correlation $Q$
     ($b_Q$, symbols) via equation (\ref{eq:b1c2_q3auto})
   		using triangles with $r_{13}/r_{12} = 2$ as a function
                of $r_{12}$. This is compared with $b_1$ derived from the ratio of
                dark matter and halo two-point correlations $\xi$ ($b_\xi$, lines) from Fig. \ref{fig:b_from_2pc}.
		Different colours denote results for the  mass samples M0 to M3 (from bottom to top) with the same colour coding as in Fig. \ref{fig:b_from_2pc}.
 		\emph{Bottom}: relative difference between $b_Q$ and $b_\xi$.
		Results for redshift $z=0.0$ are shown in the Appendix.}
   \label{fig:3pc3}
\end{figure}

In order to use the bias parameter $b_Q$ to measure the growth factor via equation (\ref{eq:D_from_2pc})
we first need to quantify deviations between $b_{\xi}$ and $b_Q$, i.e. the
linear bias $b_1$ inferred from the two-point function and the one from $Q$ in
the fit to  equation (\ref{eq:b1c2_q3auto}). If the local bias model
approximation works well, then we would expect $b_Q \simeq b_{\xi}$.
A comparison is shown for different triangle scales and mass ranges in Fig. \ref{fig:3pc3}.
In the top panel we show the linear bias derived with $\xi$ and $Q$ at
redshift $z=0.5$ as lines and symbols respectively. 

In bottom panel of Fig. \ref{fig:3pc3} we see that $b_Q$ is up to $30\%$ higher than $b_{\xi}$ at large scales,
while differences increase for smaller scales and larger values of $b_{\xi}$. 
Such deviations between $b_\xi$ and $b_Q$ have also been reported by, e.g.
\cite{M&G11}, \cite{Pollack12}, \cite{baldauf12}, \cite{chan12}, \cite{moresco_2014}. Furthermore we find
that $b_Q$ for M3 is under predicted at small scales in contrast to results for the lower mass samples. 
Deviations for small triangle sizes indicate
departures from the leading order perturbative expansion
in which equation (\ref{eq:b1c2_q3auto}) is valid,
while the strong deviations for the sample M3 suggest that the quadratic expansion of the bias function might
not be sufficient for highly biased samples. Furthermore, differences between $b_{\xi}$ and $b_Q$ are expected due to non-local contributions to the bias function,
as it has been shown in $k$-space by \citet{chan12}.
Performing the same analysis at redshift $z = 0.0$ gives very similar results, which are shown in Fig. \ref{fig:bq_covnocov} of the appendix.
We find in that case slightly larger deviations at small scales presumably due to a higher impact of non-linearities on the measurement. The
overestimations at large scales are  slightly smaller possibly as a result of smaller bias values at low redshift.
We will show in a second paper that  deviations between $b_\xi$ and $b_Q$ decrease, when galaxy-matter-matter cross-correlations
instead of galaxy-galaxy-galaxy auto-correlations are analysed. In the following we will focus on the results for $r_{12}=24$ $h^{-1}$ Mpc
which is a compromise between having small errors and sufficiently large scales for linear bias estimation.

Despite the discrepancies between  $b_Q$ and  $b_{\xi}$ shown
in Fig. \ref{fig:3pc3} we will still be able to obtain a good approximation for the growth factor $D(z)$
if $b_Q$ and  $b_{\xi}$ are related by the same multiplicative constant at different redshifts.
This is because $D(z)$ only depends on the bias ratio, as shown in equation (\ref{eq:D_from_2pc}).

\subsubsection{Bias ratio $\hat{b}$ from $Q_g$ at different redshifts}\label{bhatQ}

A fundamental limitation for the growth factor measurement described in Section \ref{sec:Dfrom2pc} is its dependence on the dark matter correlations, which
cannot be directly observed. This problem is usually tackled by employing predictions for the dark matter correlations from N-body simulations
or perturbation theory \citep[see e.g.][]{Verde, mcbride11, marin13}. Alternatively weak lensing signals
can be used as a direct probe of the total matter field \citep{Jullo12, simon13}. Both approaches can add uncertainties
and systematic effects to the galaxy bias measurement and will therefore affect constrains of cosmological parameters
derived from the growth factor.

We therefore introduce a new approach for measuring the growth factor based on the following consideration:
in equation (\ref{eq:D_from_2pc}) we see that for measuring the growth factor $D(z)$ we only require knowledge 
about the ratio of the linear bias parameters at
the redshifts $z_0$ and $z$, while the absolute bias values are
irrelevant. With the three-point correlation function we can
measure this ratio directly from the distribution of galaxies without knowing $Q_{dm}$.
We can write equation (\ref{eq:b1c2_q3auto}) for the two
redshifts $z_0$ and $z$ and combine them via $Q_{dm}$ under the assumption
that $Q_{dm}$ is independent of redshift, as shown in
Fig. \ref{fig:3pc1}. We find

\begin{equation}
	Q_g(z) = \frac{1}{\hat{b}_Q} [Q_g(z_0) + \hat{c}_Q],
	\label{eq:bD_q3auto}
\end{equation}
where we have defined $\hat{b}_Q \equiv b_Q(z) / b_Q(z_0)$ and $\hat{c}_Q = \left[c_Q(z) - c_Q(z_0)\right]/b_Q(z_0)$.
Equation (\ref{eq:bD_q3auto}) allows us to estimate the bias ratio
$\hat{b}_Q$ from $Q_g$ measurements at two different redshifts.
The measurement of  $\hat{b}$ can then be used in equation
 (\ref{eq:D_from_2pc}) to estimate $D(z)$ from  the measured $D_g(z)$.
 The results will be shown later in Section \ref{d_comparison}.

%%%%%%%%%%%%%%%%%%%%%%%%%%%%%%%%%%%%%
%                              Sub-Section V                           %
%%%%%%%%%%%%%%%%%%%%%%%%%%%%%%%%%%%%%

\subsection{Bias  $b_\tau$ from third-order moments $C_{12}$ and $S_3$}\label{sec:BM12}

Here we are interested in the joint one- and two-point third-order cumulant
moments  taken at the locations ${\bf r}_1$  and ${\bf r}_2$. 
We will estimate the first and second-order biasing coefficients 
by combining the skewness $S_3$ and reduced correlator $C_{12}$.
The skewness $S_3$ is the ratio of the one-point third-order cumulant, 
$\langle\delta^3\rangle$, and the one-point variance, $\sigma^2 \equiv \langle\delta^2\rangle$,
squared:
\begin{equation}
S_3 \equiv \frac{\langle \delta^3\rangle}{\langle\delta^2\rangle^2} \equiv
\frac{\langle \delta^3\rangle}{\sigma^4},
\end{equation}
e.g. see equation (24) in \citet{g86}.
The reduced correlator $C_{12}$  \citep{bernardeau96} is
 defined as the ratio between the joint two-point 3rd order cumulant, $ \langle\delta_1{\delta_2}^2\rangle$
over the product of the variance $\langle\delta^2\rangle$ and the two-point correlation function $\langle\delta_1\delta_2\rangle$:

\begin{equation}
C_{12}(r_{12})\equiv \frac{\langle\delta_1{\delta_2}^2\rangle}{ \langle\delta^2\rangle ~\langle\delta_1\delta_2\rangle} \equiv
\frac{\langle\delta_1{\delta_2}^2\rangle}{ \sigma^2 \xi_{12} }.
\label{c12}
\end{equation}
Note that, due to the same isotropic property as for the two-point correlation function, $C_{12}$ depends on the separation $r_{12}$ and not
on the shape of the over density. The same happens for $S_3$, which is
a spherical average over some fix smoothing radius $R$. 
Both, the skewness $S_3$ and the correlator $C_{12}$, can be seen as
two different collapsed forms of the reduced smoothed three-point correlation function $Q(r_{12},r_{13},\alpha)$,
 i.e. $S_3=3Q(0,0,0)$ and  $C_{12}(r_{12})=Q(r_{12},r_{12},0)(2+\xi_{12}/\sigma^2)$.

\subsubsection{Non-linear bias}
\label{subsec:nlbias}

Since it has been shown \citep{FG, bm} that the local non-linear bias model conserves the hierarchical properties of both, cumulants and correlators of
matter, one can express such quantities for any biased tracers (haloes or galaxies) with respect to the linear and quadratic bias coefficients
\begin{equation}
S_{3,g}  \simeq  \frac{1}{b_1}(S_{3,dm} + 3c_2)
 \label{s3h}
\end{equation}
\begin{equation}
C_{12,g}  \simeq \frac{1}{b_1}( C_{12,dm}+ 2c_2) .
 \label{c12h}
\end{equation}
Following an orginal idea of \citet{Szapudi98}, \citet{bm} worked out the explicit expressions of the bias coefficients up to fourth order.
Since in the present paper we focus on the quadratic biasing model we recall the expressions they obtained at second order.
By combining equations (\ref{c12h}) and (\ref{s3h}) one can find
\begin{equation}
b_{\tau} \equiv \frac{3 \ C_{12,dm}-2 \ S_{3,dm}} {3 \ C_{12,g}-2 \ S_{3,g}} \equiv \frac{{\tau_{dm} }}{\tau_g},
 \label{biaslin}
\end{equation}
\begin{equation}
c_{\tau} \equiv \frac{C_{12,dm} \ S_{3,g} - C_{12,g} \ S_{3,dm}}{\tau_g}.
 \label{biasquad}
\end{equation}
As in the case of $Q$ we interpret the parameters $b_{\tau}$ and $c_{\tau}$ as estimators of the first and second-order bias coefficients $b_1$ and $c_2$
respectively, while we expect this interpretation to be valid only in
the perturbation theory regime.

In practice, the skewness $S_3$ and the reduced correlator $C_{12}$ can be estimated once the density fluctuations of haloes and matter ($\delta_g$ and $\delta_{dm}$
respectively) have been smoothed on a scale $R$. In order to simplify the interfacing with theoretical predictions it is common to use a spherical Top-Hat window to smooth fluctuations.
This is done by the count-in-cell estimators for the discrete $S_{3,N}$ and $C_{12,N}$,
which are described in \citet{bm} and used in this analysis. We correct these estimations from shot noise by assuming the local Poisson process approximation \citep{Layser}. 
Note that, in order to be able to handle the large number of dark matter particles, we use only $1/700$ of the total number of particles
in the dark matter simulation output. This introduces additional shot noise errors but we have tested that it does not affect the measurements.
To measure $S_{3,N}$, we set up a regular grid of spherical cells. From the number of dark matter particles per cell we derive the corresponding
number count fluctuations ($\delta_N$), which we use to estimate the skewness via
$$
S_3=\frac{\langle\delta_N^3\rangle-3\langle\delta_N^2\rangle\bar N^{-1}+2\bar N^{-2}}{ (\langle\delta_N^2\rangle-\bar N^{-1})^2},
$$
where $\bar N$ is the number of particles per grid cell, averaged over the volume of the simulation box.

The reduced correlator, $C_{12}$, is measured via the two-point count-in-cell estimator \citep{bm}. We therefore set up a regular grid of spherical cells, which are separated by two times the smoothing radius $R$
(hereafter referred to as seeds) and place an isotropic motif of spheres around each seed at separation $r_{12}$.
The two-point moments of the density field are then measured as the correlation between density fluctuations in the spheres of the motif and those in the seeds. This method allows for measurements of
two-point statistics, even in the low separation limit when the spheres of the motif touch the spheres of the seeds, without being affected by any choice of  distance bins. As for the skewness, we correct
for shot noise assuming a Poisson sampling, which leads to
$$
C_{12}=\frac{ \langle\delta_{N,1}\delta_{N,2}^2\rangle-2\langle\delta_{N,1}\delta_{N,2}\rangle\bar N^{-1} }{  \langle\delta_{N,1}\delta_{N,2}\rangle(\langle\delta_N^2\rangle-\bar N^{-1}) }.
$$
Once we have measured the skewness and the reduced correlator, the linear and quadratic bias $b_\tau$ and $c_\tau$ can be estimated for any chosen smoothing radius $R$ and for any
ratio $n\equiv r_{12}/R$ with equation (\ref{biasquad}). Errors of the measurements are computed by Jackknife resampling,  as described in Section \ref{sec:err}.

A theoretical prediction for the skewness and the reduced correlator can be derived with perturbation theory \citep[PT,][]{bernardeau92,bernardeau96}:
\begin{equation}
S_{3}^{PT}   =  \displaystyle   \frac{34}{7} + \gamma_R     \label{s3pt}, \\
\end{equation}
\begin{equation}
C_{12}^{PT}(r_{12})  =  \displaystyle    \frac{68}{21} + \frac{\gamma_R}{3} + \frac{\beta_R(r_{12})}{3},
\label{c12pt}
\end{equation}
where $\gamma_R\equiv d\ln\sigma_R^2/d\ln R$ and $\beta_R(r) \equiv d\ln \xi_R(r_{12})/d\ln R$ are, respectively, the logarithmic derivatives of the variance and the two-point correlation function of the smoothed
field of density fluctuations $\delta_R$ with respect to the smoothing radius $R$. Note that expression (\ref{c12pt}) has been obtained in the large separation limit ($r_{12} \ge 3R$). 
\begin{figure}
\includegraphics[width=85mm,angle=0]{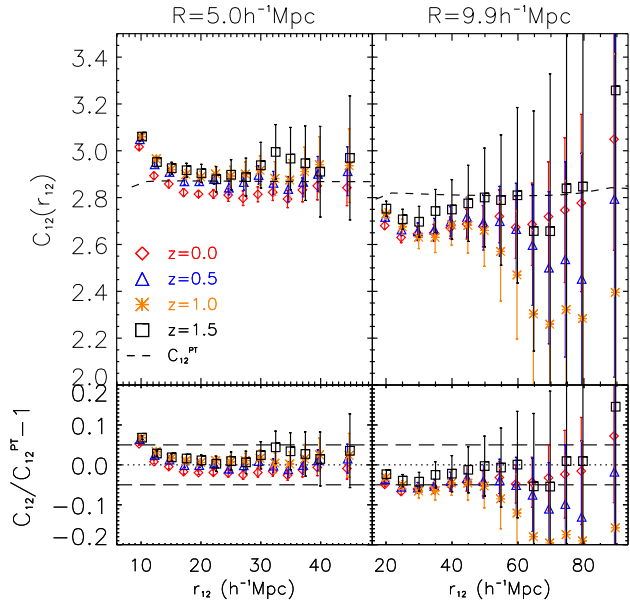}
\caption{\small   {\it Top}:  reduced correlator $C_{12}$ of the MICE-GC dark matter field as a function of the separation $r_{12}$ measured in the comoving outputs
at redshifts $0.0$, $0.5$, $1.0$ and $1.5$ (diamonds, triangles, crosses and squares respectively) compared to the perturbation theory (PT) prediction using a linear power spectrum
(dashed line) with smoothing radii $R=5.0$ $h^{-1}$Mpc (left) and $R=9.9$ $h^{-1}$Mpc (right).  {\it Bottom}: relative difference between measurements and PT  prediction.
Black dashed lines denote $\pm 5$\% deviation.}
\label{c12_r12}
\end{figure}

Fig. \ref{c12_r12} gives an insight into the accuracy of the theoretical prediction of the reduced correlator $C_{12}$ from equation (\ref{c12pt}) for two different smoothing radii $R$.
The results, shown in Fig. \ref{c12_r12} also show that on a wide range of separations ($10$-$60$ $h^{-1}$Mpc) estimating the correlator at different epochs has no significant impact on the
measured value, as predicted by perturbation theory. In fact, the
values estimated at the four considered simulation snapshots ($z=0$,
$z=0.5$, $z=1.0$ and $z=1.5$) vary by less than $5$\% (bottom panel).

The agreement with perturbation theory requires two important
ingredients. First that the separation is much larger than the smoothing scale.
Second, that we include the $\beta_R$ term in the
prediction \citep{bm}. Previously, this term was considered negligible 
\citep{bernardeau96,GFC} and this resulted in a mismatch with
numerical simulations, attributed to non-linear effects in the
spherical collapse model \citep{GFC}. 

\begin{figure}
\includegraphics[width=85mm,angle=0]{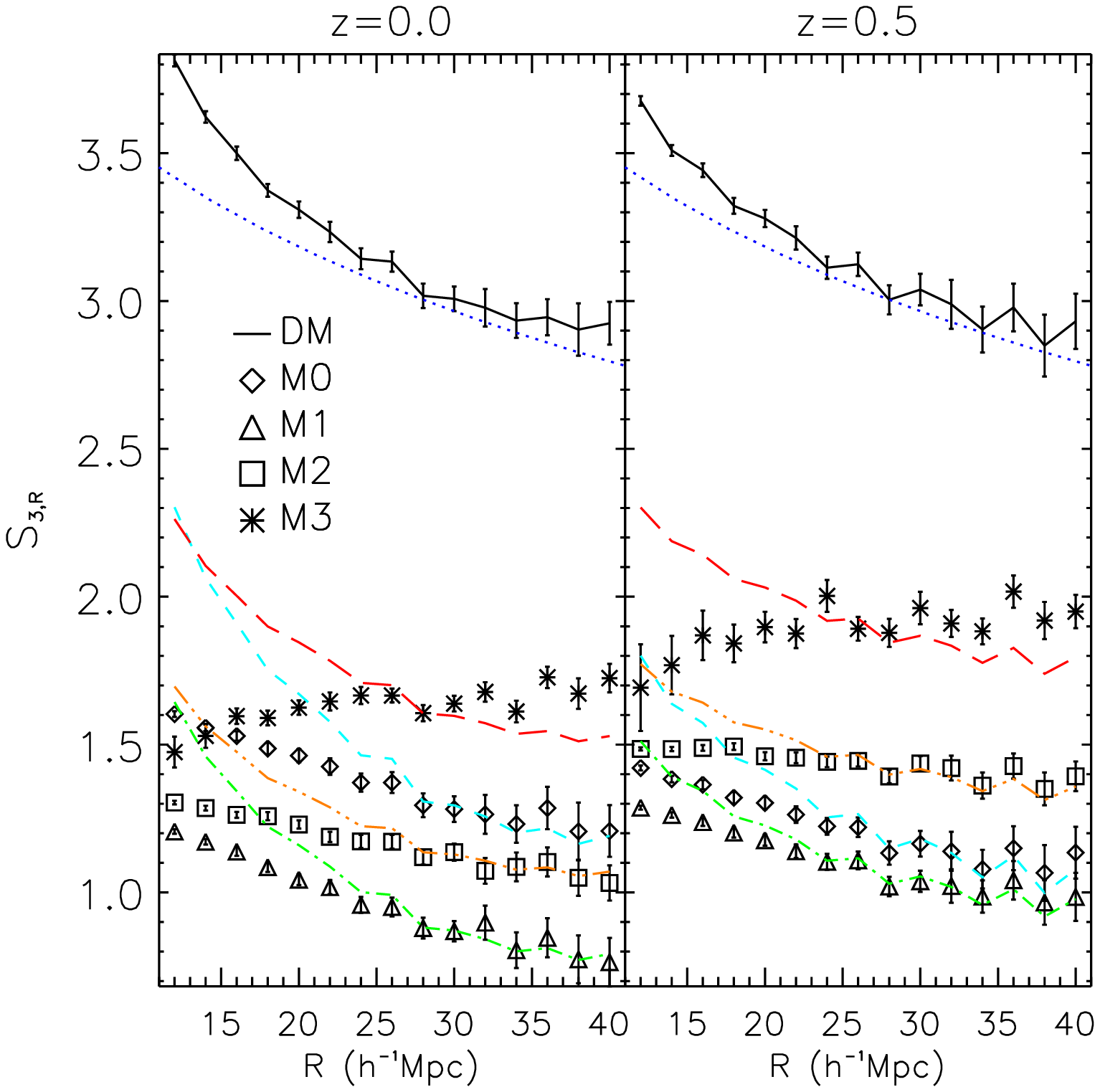}
\includegraphics[width=85mm,angle=0]{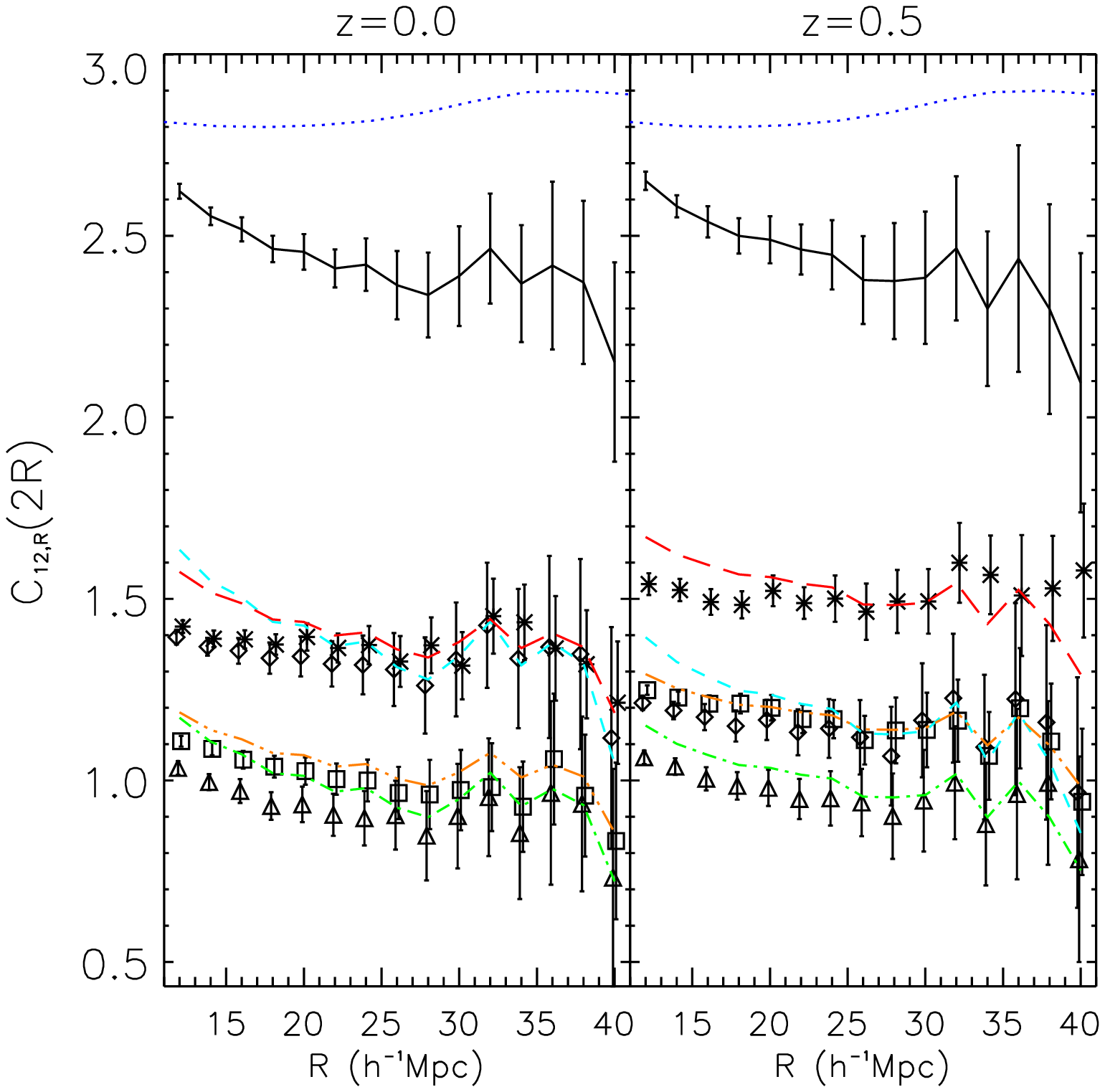}
\caption{Skewness $S_3$ and reduced correlator $C_{12}$ (top and bottom panel respectively) 
measured from the dark matter field (solid line) and the four halo mass samples M0-M3 (symbols) in the MICE-GC comoving outputs at redshifts
$z=0.0$ and $z=0.5$ (left and right respectively) as a function of smoothing radius $R$.  The blue dotted lines display the tree level perturbation theory predictions for $S_3$ and $C_{12}$
respectively given by equations (\ref{s3pt}) and (\ref{c12pt}). Coloured lines denote $S_{3,g}$ and $C_{12,g}$ expected from equations (\ref{s3h}) and (\ref{c12h}).}
\label{s3c12}
\end{figure}
To increase the statistical signal of our measurements, we set the correlation length to be twice the smoothing radius, i.e. $r_{12}=2 R$.
The cells, used for the $C_{12,g}$ measurements, are consequently positioned side by side.

The Jackknife errors of $C_{12}$ from the dark matter field increase with redshift. This can be explained by shot noise being redshift independent as the number of particles is conserved,
while the amplitude of the correlation decreases with redshift, causing smaller signal-to-noise ratios.
Furthermore, one could expect that shot noise has a higher impact on $C_{12}$ at small smoothing scales since the smoothing window encloses on average less particles. However, we
observe the opposite trend, presumably because a larger smoothing scale implies a smaller number of independent measurements due to the finite comoving volume of the simulation.
This also explains the increase of errors with separation $r_{12}$.

$S_3$ and $C_{12}$ measurements of the MICE-GC dark matter field are presented in Fig. \ref{s3c12}, where they are contrasted with the corresponding quantities measured for haloes.
At large smoothing scales the $S_3$ measurements for dark matter are in good agreement with the prediction from equation (\ref{s3pt}), represented by the blue dotted line.
At small smoothing radii ($R<20$ $h^{-1}$Mpc) we find the measurements to be significantly higher than the predictions. For haloes we can see that $S_3$ does not
increase monotonically with mass. In fact, at $R=30$ $h^{-1}$Mpc and $z=0$, for the low mass sample M0 the skewness is around $1.3$, then drops
down to $0.8$ for M1, increases to $1.2$ for M2 and finally reaches the value $1.6$ for the high mass sample M3. A similar tendency is observed at higher redshift. This non-monotonic behavior
is qualitatively expected by the spherical collapse \citep{Mo97} and ellipsoidal collapse \citep{Sheth01} predictions and is in quantitative agreement with the measurements of \citet{AB&L08},
performed in a different simulation with a lower mass resolution and on a smaller mass range than the one reachable with the MICE-GC simulation. 

\begin{figure*}
\centerline{\hskip 2.5cm \includegraphics[width=200mm,angle=0]{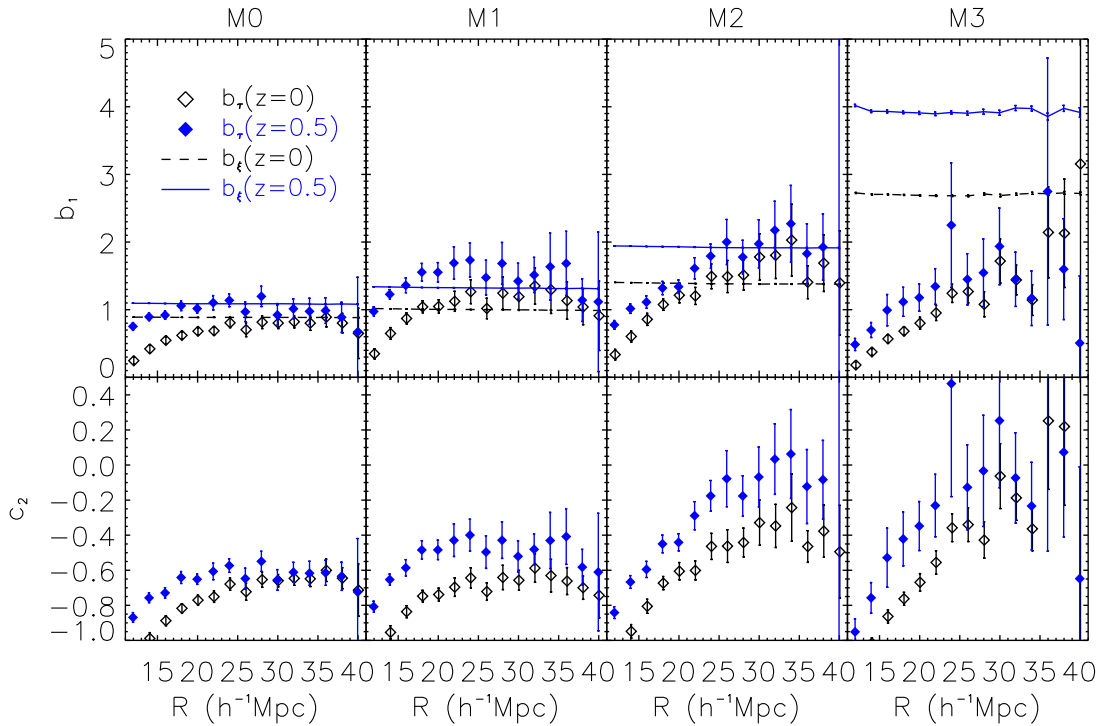}}
\caption{\small {\it Top:} linear bias parameter $b_1$ obtained from the $\tau$ estimator ($b_\tau$) via equation (\ref{biaslin})  (diamonds) for various smoothing radii $R$
compared to the reference linear bias obtained from the two-point correlation $\xi$ via equation (\ref{bxi}) ($b_\xi$, lines) at the redshifts $z=0.0$ (black) and $z=0.5$ (blue).
{\it Bottom:} quadratic bias parameter $c_\tau$ estimated 
from equation (\ref{biasquad})  (diamonds) with same colour coding as in the top panel. Columns show results for the mass samples M0-M3.}
\label{b1c2mz}
\end{figure*}

The bottom panel of Fig. \ref{s3c12} shows for the first time how the mass of the chosen haloes affects the reduced correlator, $C_{12}$. Comparing the effect of biasing on the shape of
$S_3$ with its effect on $C_{12}$, one can see that both follow the local bias model for large smoothing radii $R$. The local bias model seems to be less accurate for $S_3$ as a function of $R$,
since its shape is systematically affected in a non-linear way. Moreover, we find the shape modification to increase with halo mass.
Despite the fact that the predictions from equation (\ref{c12pt}) reproduce the large scale behavior of $C_{12}$ for haloes, we confirm significant deviations from the dark matter
measurements at small separation ($r_{12}=2R$), even after taking into account the $\beta$-term. But note that equation (\ref{c12pt}) has been obtained in the large separation limit ($r_{12}>>R$),
so this is not totally unexpected. However, one can see in Fig. \ref{c12_r12} that for larger separations the theory is in remarkably good agreement with measurements
even for small smoothing radii such as $R=5$ $h^{-1}$Mpc. The local bias model works for $C_{12}$ if we use the matter measurements as input,
despite their disagreement with perturbation theory. This is because the local model is an expansion on $\delta$ but does not require $r_{12}>>R$.

Turning to bias estimators, using equations (\ref{biaslin}) and (\ref{biasquad}) we
measured $b_\tau$ and $c_\tau$  in each mass sample at both redshifts
with respect to the smoothing radius $R$. Measurements are
displayed in Fig. \ref{b1c2mz}, together with the bias estimator
$b_\xi$ previously described. Estimators in equation (\ref{biaslin})
for $b_1$ and $c_2$ exhibit a significant scale dependency before
converging to a constant value. This allows us to set up effective
scale ranges in the fitting procedure used to measure the linear and
quadratic bias. Comparing the scale dependency obtained for
the various mass bins and redshifts we conclude that above
$26$ $h^{-1}$Mpc both $b_{\tau}$ and $c_{\tau}$ are independent from the
considered smoothing scale. We therefore measure them by performing a
fit between $R=26$ and $40$ $h^{-1}$Mpc. As we showed that the shape of
$S_3$ is highly affected at small scales, in particular at high halo
masses, we can reasonably conclude that, on one hand, the scale dependency of
$b_\tau$ and $c_\tau$ results from the skewness of haloes
and, on the other hand, the large discrepancy observed between $b_\xi$ and $b_\tau$ (right panel of Fig.~\ref{b1c2mz})
is due to an underestimation of the skewness for massive haloes (we discuss this effect in Section \ref{b_comparison}). 

\subsubsection{Bias ratio $\hat{b}$ from $\tau_g$ at different redshifts}

In Subsection \ref{bhatQ} we introduced a way of estimating the bias ratio $\hat b$ using $Q_g$ to measure the linear growth of structures directly from galaxies
(or  haloes), i.e. without assuming any modelling of the power spectrum nor of the bispectrum of matter fluctuations. Following the same idea we show
now that we can measure $\hat b$ using the $b_\tau$ estimator from equation (\ref{biaslin}) at the two redshifts: $z_0=0$ and $z$.
Assuming that $\tau_{dm}$ does not depend on redshift,  it is straightforward to show that:

\begin{equation}
\hat b_\tau \equiv \frac{b_\tau(z)}{b_\tau(z_0)} = \frac{\tau_g(z_0)}{\tau_g(z)}.
\label{bgrow}
\end{equation}
The assumption that $\tau_{dm} \equiv 3\,C_{{12,dm}}-2\,S_{{3,dm}} $ does not evolve with redshift is strongly supported by theory and simulations.
Both $S_3$ and $C_{12}$ are expected to be weakly sensitive to
redshift in perturbation theory. Measurements, shown in Fig. \ref{c12_r12}, illustrate
that  $C_{12,dm}$ is weakly affected by redshift evolution, while results in Fig.~(\ref{s3c12}) show that,
on large scales ($> 20$ $h^{-1}$Mpc), the skewness $S_{3,dm}$ does not present significant redshift dependency
(see also \citet{bernardeau02} and references therein). 

%%%%%%%%%%%%%%%%%%%%%%%%%%%%%%%%%%%%%
%                              Sub-Section VI                          					%
%%%%%%%%%%%%%%%%%%%%%%%%%%%%%%%%%%%%%

\subsection{Errors estimation and fitting}\label{sec:err}

Since we use either one simulation at various comoving outputs ($z=0$,
$z=0.5$, $z=1.$ and $z=1.5$), or one light cone, we estimate the errors of
$\xi$, $Q$, $S_3$, $C_{12}$, $b_{\xi}$, $b_Q$, $b_\tau$, $c_Q$, $c_\tau$ and $D$ measurements
by Jackknife resampling. The Jackknife samples of the complete
comoving output are constructed from $64$ cubical sub-volumes while in case of
the light cone we use $100$ angular regions (with equal volume at each
redshift bin) in right ascension and declination on the sky. Following \cite{norberg09},
we generate for any statistical quantity $X$ a set of pseudo-independent measurements
($X_j$), from which we compute the standard deviation $\sigma_X$ around the mean $\bar X$ (computed
on the complete volume) as

\begin{equation}
\sigma_X=\sqrt{\frac{(n-1)}{n}\sum_{j=0}^{n}(X_j-\bar X)^2},
\label{jk}
\end{equation}
where $n$ is the number of Jackknife samples.

For all three bias estimations $b_\xi$, $b_Q$ or  $b_\tau$, we use the same fitting procedure, which takes into account the covariance between $\xi$, $Q$ and $\tau$
measurements at different separations, opening angles and smoothing scales ($r_{12}$, $\alpha$, $R$ respectively). The covariance matrix $C$ is computed from the deviation
matrix  $A$, which in turn is estimated by Jackknife resampling as well: a measurement in the $j^{th}$ Jackknife sub-volume and for the $i^{th}$ separation, angle or scale is written $X_{ij}$.
Each element $A_{ij}$ of the deviation matrix is calculated as $A_{ij}=X_{ij}-\bar X_i$. Again the mean $\bar X_i$ is the measurement on the complete volume.
The covariance matrix can then be computed straightforwardly 

\begin{equation}
C=\frac{n-1}{n}A^TA. 
\label{eq:cova}
\end{equation}
The bias ratio $\hat{b}_\tau$ can be fitted from results at different smoothing scales in a very simple way, because the first- and second-order bias coefficients can be estimated
separately in the $\tau$ formalism (see equation (\ref{bgrow})). 
Deriving $\hat{b}_Q$ is more complicated as it requires a two-parameter fit,
due to  the mixing of the bias coefficients (see equation (\ref{eq:bD_q3auto})).
The main problem arises from the fact that at a given redshift the errors of $Q_g$ are correlated between the various angles.
Furthermore the reduced three-point correlation can also be correlated between the two redshifts $z_0$ and $z$, where $z_0$ is the reference redshift.
Based on equation (\ref{eq:bD_q3auto}), we define the variable
\begin{equation}
Z \equiv Q_g(z_0) - (\hat b Q_g(z) + \hat c),
\label{defz}
\end{equation}
and vary $\hat b$ and $\hat c$ in order to obtain $Z=0$ for all angles $\alpha$. In other words we want to measure the posterior probability distribution (hereafter referred to as likelihood
$L(\hat b, \hat c)$) of the two parameters $\hat b$ and $\hat c$ given that $Z$ is expected to be $Null$.
Assuming a multivariate normal distribution of $Z$, one can write the log-likelihood $\mathcal{L}\equiv -2\ln(L)$ as for measuring a given $Z$

\begin{equation}
\mathcal{L}=B + \ln(|C_Z|) + \chi^2,
\label{loglik}
\end{equation}
where $C_Z$ is the covariance matrix of the $Z$, $B$ is a normalisation constant and $\chi^2\equiv \sum_{i,j} Z_jC_{Z,ij}^{-1}Z_i$. Note that, if the covariance matrix does not depend on the parameters
of the model, then the second term in expression (\ref{loglik}) can be absorbed in the normalisation constant $B$. However, from definition (\ref{defz}) follows that $C_Z$ explicitly depends on the
fitting parameters $\hat b$ and $\hat c$. It can therefore be obtained from the covariance matrix of $Q_g(z_0)$, $Q_g(z_j)$ and from the cross-covariance of $Q_g(z_0)$ and $Q_g(z_j)$: 

\begin{equation}
C_Z=C_X + \hat b^2 C_Y - \hat b\left ( C_{XY} + C_{XY}^{\intercal} \right ),
\label{cov}
\end{equation}
which explicitly shows the dependency of the covariance matrix $C_Z$ on the fitting parameter $\hat b$. Note that $C_X$ and $C_Y$
are respectively the covariance matrix of  $Q_g(z_0)$ and $Q_g(z_j)$ computed with equation (\ref{eq:cova}). The cross-covariance
matrix $C_{XY}$ is defined as

\begin{equation}
C_{XY,ij}=\frac{n-1}{n}(X_j-\bar X)(Y_i-\bar Y),
\label{eq:covaX}
\end{equation}
where $n$ is the number of elements in both $X$ and $Y$. In practice we shall neglect the correlation between redshift bins, so that $C_{XY}=C_{XY}^{\intercal}=0$ in equation (\ref{cov}). 
Otherwise the inverse covariance matrix $C_{XY,ij}^{-1}$ had to be computed for each tested value of  $\hat b$.
The estimate of $\hat b$ and its error are obtained by marginalising over the $\hat c$ parameter via the posterior marginalised log-likelihood
$$
\mathcal{L}(\hat b)=-2\ln\left\lbrace \int L(\hat b, \hat c)d\hat c \right\rbrace.
$$ 
Testing the assumption that measurements at different redshifts are uncorrelated, we verified that the correlation coefficient remains very small compared to unity. It follows that the square
of the relative error for the bias ratio is obtained by summing in quadrature the relative errors of $b_X(z_j)$ and $b_X(z_0)$. Then, since $b_Q$ and $b_\tau$ are third-order estimators,
we checked that the error of the halo growth factor $D_h$ is negligible with respect to the error obtained for the bias ratio $\hat b$.

%%%%%%%%%%%%%%%%%%%%%%%%%%%%%%%%%%%%%%%%%%%%%%%%%%%%%%%%%%%%%
%%%%%%%%%%%%%%%%%%%%%%%%%%%%%%%%%%%%%%%%%%%%%%%%%%%%%%%%%%%%%
%%                                                RESULTS (Section IV)                                                %%
%%%%%%%%%%%%%%%%%%%%%%%%%%%%%%%%%%%%%%%%%%%%%%%%%%%%%%%%%%%%%
%%%%%%%%%%%%%%%%%%%%%%%%%%%%%%%%%%%%%%%%%%%%%%%%%%%%%%%%%%%%%

\section{Results}\label{sec:results}

As we have pointed out in the sections \ref{sec:Dfrom2pc}-\ref{sec:BM12} we use growth independent bias measurements from third-order statistics to break the growth-bias degeneracy that appears in
growth measurements from two-point correlations. This approach is limited by the accuracy and the precision with which third-order statistics can measure galaxy bias.
We study the differences between bias from second- and third-order correlations for different redshifts and halo mass ranges and present the results in Section \ref{b_comparison}. In Section \ref{d_comparison} we show the resulting estimations for the linear growth measurements.

Alternatively to the direct approach of growth measurement described above, we have introduced a new method which does not require any modelling of third-order clustering of dark matter.
It takes advantage of the fact that only the ratio of the bias parameters at two redshifts needs to be known to break the growth-bias degeneracy.
This bias ratio can be directly measured from third-order statistics of the halo field (see Section  \ref{sec:Dfrom2pc}-\ref{sec:BM12}). 
In Section \ref{d_comparison2} we compare growth factor measurements from our new method and the more  common method of combining
second- and third-order statistics with theoretical predictions (or simulations) for the dark matter field.
In Section \ref{f_comparison} we present growth rate measurements derived with and without third-order correlations of dark matter.

%%%%%%%%%%%%%%%%%%%%%%%%%%%%%%%%%%%%%
%                              Sub-Section I                           %
%%%%%%%%%%%%%%%%%%%%%%%%%%%%%%%%%%%%%

\subsection{Bias comparison}\label{b_comparison}

\subsubsection{Measurements in the comoving outputs}

\begin{figure*}
\centerline{\includegraphics[width=120mm,angle=0]{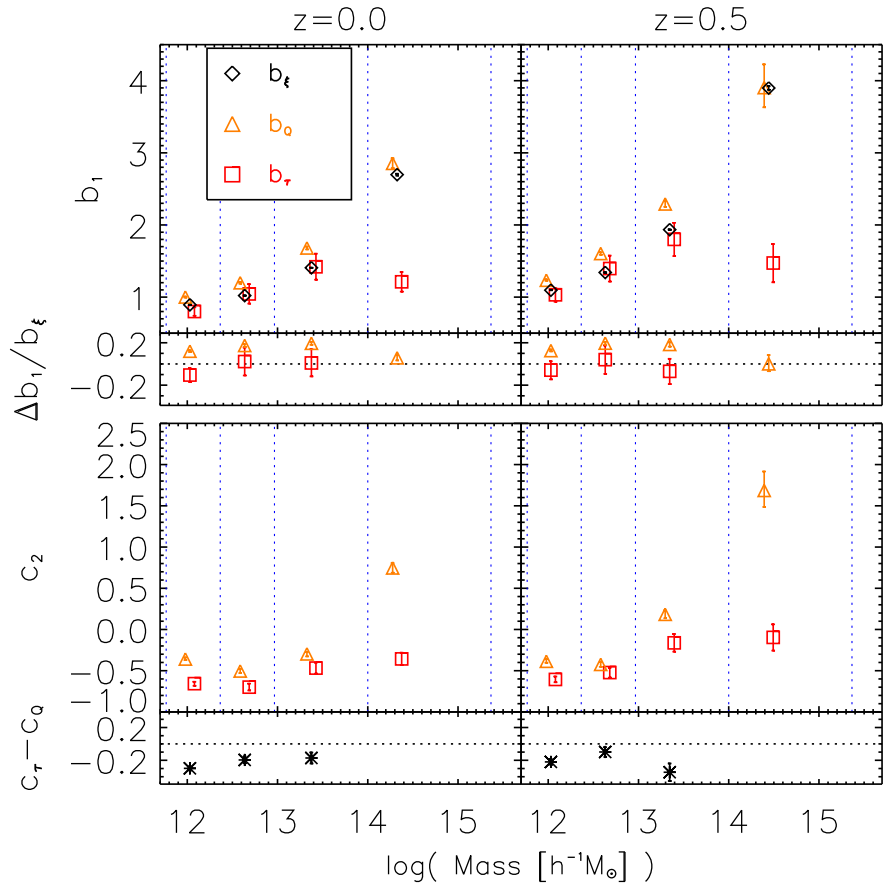}}
\caption{\small {\it Top}:
linear bias parameters $b_1$ measured  with
two-point correlations  $\xi$ via equation (\ref{bxi}) ($b_\xi$, black diamonds),
reduced three-point correlations $Q(24,48,\alpha)$ via equation (\ref{eq:b1c2_q3auto}) ($b_Q$, orange triangles) and the $\tau$ estimator from
the third-order moments $S_3$ and $C_{12}$ via equation (\ref{biaslin}) ($b_\tau$, red squares) 
in the MICE-GC comoving outputs at redshift $z=0.0$ (left) and $z=0.5$ (right) for the mass samples M0-M3.
The relative difference between $b_Q$ and $b_\tau$ with respect to $b_\xi$ is shown below.
{\it Bottom:}  quadratic bias coefficient $c_2$ estimated from $Q$ and $\tau$ via equation (\ref{eq:b1c2_q3auto}) and (\ref{biaslin}) respectively in the same
colour coding as in the top panel.
The absolute difference between $c_Q$ and $c_\tau$ is shown below.
On all panels, the range of each mass sample is delimited by the blue dotted lines. The diamonds are positioned at the average
mass of haloes contained in each mass sample and we shifted triangles and squares for clarity.}
\label{fig:results4}
\end{figure*}

In Fig. \ref{fig:results4} we show the values of the linear and quadratic bias parameters $b_1$ and $c_2$, measured with the $Q$ and $\tau$ estimators ($b_Q, c_Q$ and $b_{\tau},c_{\tau}$)
in the comoving outputs at redshift $z=0.0$ and $z=0.5$. The bias parameters from $Q$ are estimated from triangles with fixed legs of $24$ and $48$ $h^{-1}$Mpc
(see Table \ref{triangle_configs} for details) using $18$  opening angles $\alpha$ with values between 0 and 180 degree as shown in Fig. \ref{fig:3pc2}. The $\tau$
bias estimations are based on fits of $b_{\tau}$ and $c_{\tau}$ between $26  < R <  40$ $h^{-1}$Mpc using ($r_{12} = 2R$) configurations (see Fig. \ref{b1c2mz}).
All error bars denote the standard deviation derived from $64$ Jackknife samples as described in Section \ref{sec:err}.

In the same figure we compare our measurements of the linear bias parameter from third-order statistics with $b_\xi$ computed from the two-point correlation between
$20$ and $60$ $h^{-1}$Mpc by showing the absolute values as well as the relative differences.
In case of $c_2$ we show the absolute instead of the relative difference since we have no reference values from two-point correlations.

The linear and quadratic bias parameters from both estimators increase with mass and redshift, while their absolute values differ in several aspects.
As demonstrated already in Fig. \ref{fig:3pc3}, $Q$ overestimates the linear bias in all mass samples by a factor between $20-30\%$ with respect to $b_{\xi}$.
The good agreement between $b_{\xi}$ and $b_Q$ for the high mass sample M3 only appears for the chosen triangle configuration of (24,48,$\alpha$).
The overestimations confirm findings from \citet{M&G11}, \cite{Pollack12} and \cite{chan12}, while the large volume and resolution
of the MICE-GC simulation allows us to extend this bias comparison to a wider range
of masses than probed previously.  \cite{chan12} argued that the mismatch
of the linear bias by $Q$ is expected from non-local
contributions to the bias function. But note that the comparison with the results here is not direct.
Our results are in configuration space (not in Fourier space), for  halo-halo-halo correlations
(not halo-matter-matter) and for a different cosmology. Moreover,
the deviations found in \cite{chan12} depend linearly on the $b_1$
value, while we find here a shift by a factor that is roughly
independent of mass (and therefore of $b_1$).
We will explore some of these differences
in a separate paper.

On the other hand,
the $\tau$ bias estimation, $b_\tau$, shows a good agreement with $b_{\xi}$ for the lower mass bins, but at a price of larger
errors. This suggests that the $\tau$ estimator is less affected by such non-local contributions than $b_Q$,
which is also expected from theory, since the isotropy of the estimators of $S_3$ and $C_{12}$ could wash out the non-local effects. 
$\tau$ strongly underestimates the linear bias parameter for the highest masse bin (M3). This might be caused by discreteness effects
which need to be corrected in the $\tau$ estimation (note that this does not affect $Q$). Here we have used Poisson shot noise corrections,
but these are likely to be incorrect for massive haloes because of exclusion effects \cite[see e.g.][]{M&G11}.
As a result, for haloes which are too massive, the correct estimation of the skewness is far from being trivial.
Assuming a wrong shot noise correction leads to an underestimation of $S_{g,3}$, which causes $\tau_g$ to be
over-predicted and translates for the $b_\tau$ estimator into a large underestimation of the linear bias and presumably
of the second-order bias (left panel of Fig.~\ref{fig:results4}).

Besides non-local terms, the discrepancies between $b_{\xi}$, $b_{\tau}$ and $b_Q$, highlighted in Fig. \ref{fig:results4}, can be caused
by various other effects, such as stochasticity or contributions of higher-order terms to the bias expansion (equation \ref{eq:def_biasmodel}).
The Jackknife estimation of the errors and the covariance matrix introduces an additional uncertainty in the bias measurement
(see Appendix \ref{sec:app}).

The larger errors of the $b_\tau$  measurements with respect to those from $b_Q$ are a consequence of the larger errors
in $C_{12}$ and $S_3$ with respect to $Q$ in Fig. \ref{fig:3pc2} and \ref{s3c12}. The larger errors in $S_3$ and  $C_{12}$ result 
from the larger smoothing scales compared to those used to compute Q,
which leads to a lower number of independent measurements. An additional contribution to the higher scatter of the $S_3$
and $C_{12}$ results from the grid used in the  estimation. It neglects roughly $50 \%$ of the volume since the seeds 
do not overlap with each other. An additional minor source of error comes from the fact that, for practical reasons,
only $1/700$ of the total number of dark matter particles are used to measure $S_3$ and $C_{12}$ of matter,
which is discussed in Subsection \ref{subsec:nlbias}. For both $\tau$ and $Q$ errors could be improved by including
more configurations and optimal weighting.

Regarding the quadratic bias coefficient $c_2$ we found that the estimated values of $c_Q$ and $c_\tau$ are in significant disagreement for mostly all mass bins. 
We will study this results in more detail by comparing similar estimations from halo-matter-matter cross-correlation with predictions from the peak background split model
in a separate paper.

\subsubsection{Measurements in the light cone}

\begin{figure*}
\centerline{\includegraphics[width=170mm]{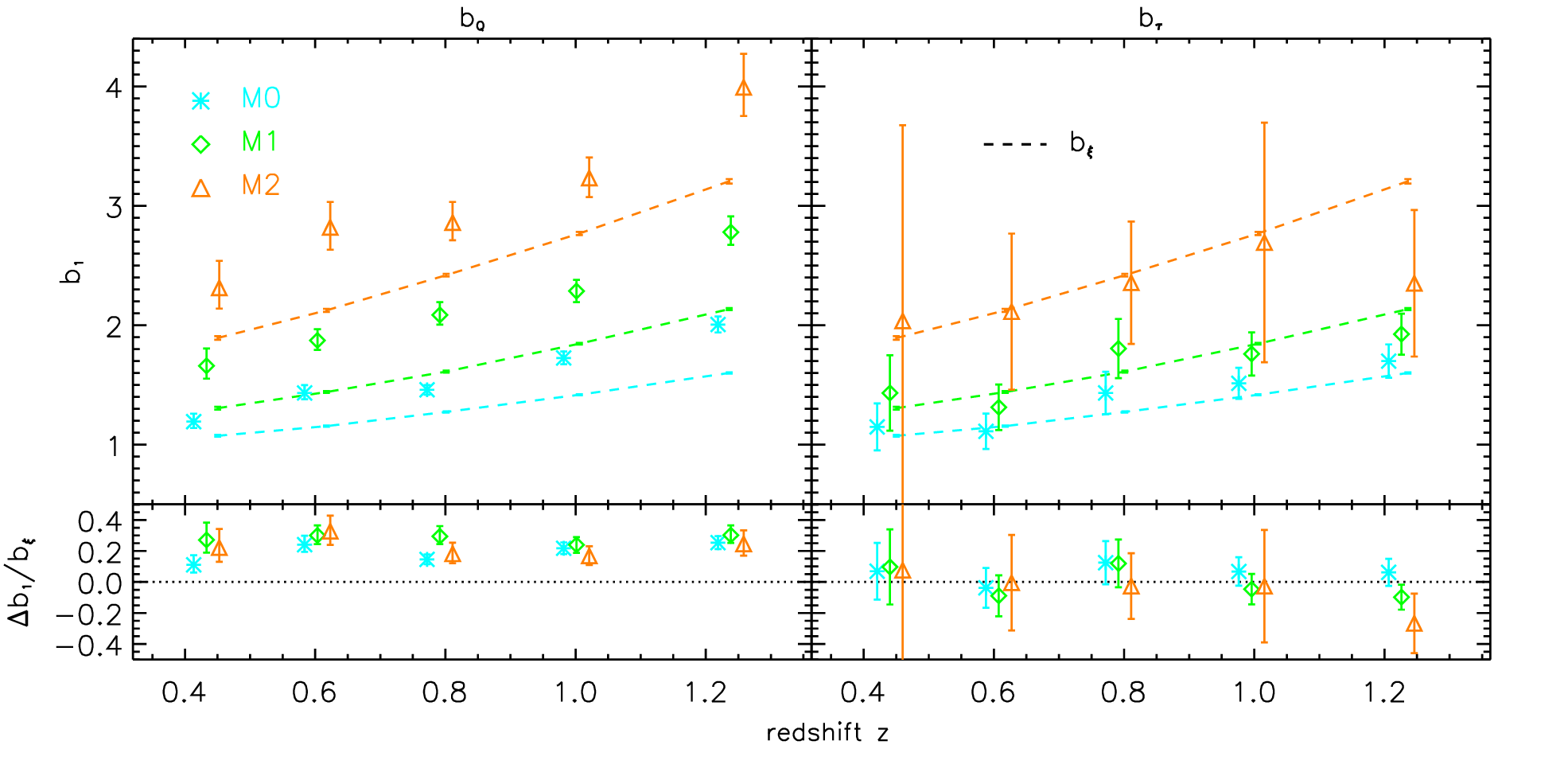}}
\caption{Linear bias as a function of the redshift estimated from $b_Q$ (left panel) and $b_\tau$ (right panel) in three mass bins of MICE-GC light cone compared
to the linear bias estimated from the two-point correlation function (coloured dashed lines).  Blue crosses, green diamonds and orange triangles correspond
respectively to M0, M1 and M2. The colour code is the same as for $b_\xi$ in Fig. \ref{fig:b_from_2pc}.}
\label{fig:results6}
\end{figure*}

To be more realistic, we conduct bias measurements in a light cone, which is constructed from the MICE-GC simulation and includes redshift evolution of structures.
The total volume probed by the light cone is about $15$ $h^{-3}$ Gpc$^{3}$ and we consider 
an octant of the sky (about 5000 deg$^2$). 
We study the deviation between the different bias estimations in five redshift bins between $0.4< z<1.42$ using the mass samples M0, M1 and M2.
We do not present results for the highest mass sample M3 and for smaller redshifts, since the results are strongly scattered due to small numbers of haloes (see Fig. \ref{fig:nofz}).
However, this mass and redshift range was previously analysed using the comoving outputs of the same simulation.

For measuring the bias we use the same (24,48,$\alpha$) configurations for $Q$ as in the comoving output. In the case of  $b_\tau$ 
we use ($r_{12} = 2R$) configurations as in the previous analysis and we perform a fit over the scale range $16$-$30$ $h^{-1}$Mpc for the mass bins M0 and M1, while we restrict this range to $25$-$30$ $h^{-1}$Mpc in case of M2.
These new fitting ranges are motivated by Fig. \ref{b1c2mz}, which shows that at the redshift $z=0.5$ the scale dependency of $b_\tau$ can be neglected on those ranges and for the corresponding mass bins. To maximize the statistical power of the estimator we now
use overlapping cells to avoid neglecting roughly half of the data in the space between the seeds used to smooth the particle distribution.
Note that in this case we estimate errors by Jackknife resampling of $100$ angular regions of the light cone (see Section \ref{sec:err}).

The results for the $b_Q$  and $b_\tau$ estimator are shown together with $b_\xi$ in Fig. \ref{fig:results6}. 
They confirm that $b_Q$ tends to overestimate the bias for the lower mass bins by about $30\%$.
Moreover it shows that the ratio between $b_\xi$ and $b_Q$ is roughly a constant with respect to the redshift or mass bins.
On the other hand  $b_\tau$ seems to be an unbiased estimate of the linear bias coefficient, while the measurements are noisier.

Since in our approach we aim at measuring the linear bias in order to extract information about the growth factor $D$ of linear matter fluctuations,
we focus on deriving a direct measurement of it in the following section.

%%%%%%%%%%%%%%%%%%%%%%%%%%%%%%%%%%%%%
%                              Sub-Section II                          %
%%%%%%%%%%%%%%%%%%%%%%%%%%%%%%%%%%%%%

\subsection{Growth Measurements}
\label{d_comparison}

In this subsection we present the growth factor $D$ and the growth rate $f$,  measured in the MICE-GC light cone via the 
equations (\ref{eq:D_from_2pc}) and (\ref{growratedis})  respectively. We obtained these measurements with the linear bias $b$, estimated
with $Q$ and $\tau$ (equation (\ref{eq:bD_q3auto}) and (\ref{biaslin}) respectively) at the redshift $z$ and the reference
redshift $z_0$. These growth measurements are compared to those from our new approach for measuring $D$ and $f$ via the bias ratio
$\hat b(z) = b(z)/b(z_0)$. We can derive $\hat b(z)$ directly by comparing  $Q$ and $\tau$ of galaxies (or haloes) at $z$ and $z_0$
(equations (\ref{eq:bD_q3auto}) and (\ref{bgrow})). This new approach allows us to measure the growth of dark matter using
only the observable second- and third-order galaxy (or halo) correlations without the corresponding dark matter statistics
(see Sections \ref{sec:3pc} and \ref{sec:BM12}).

\begin{figure*}
\centerline{\includegraphics[width=120mm,angle=270]{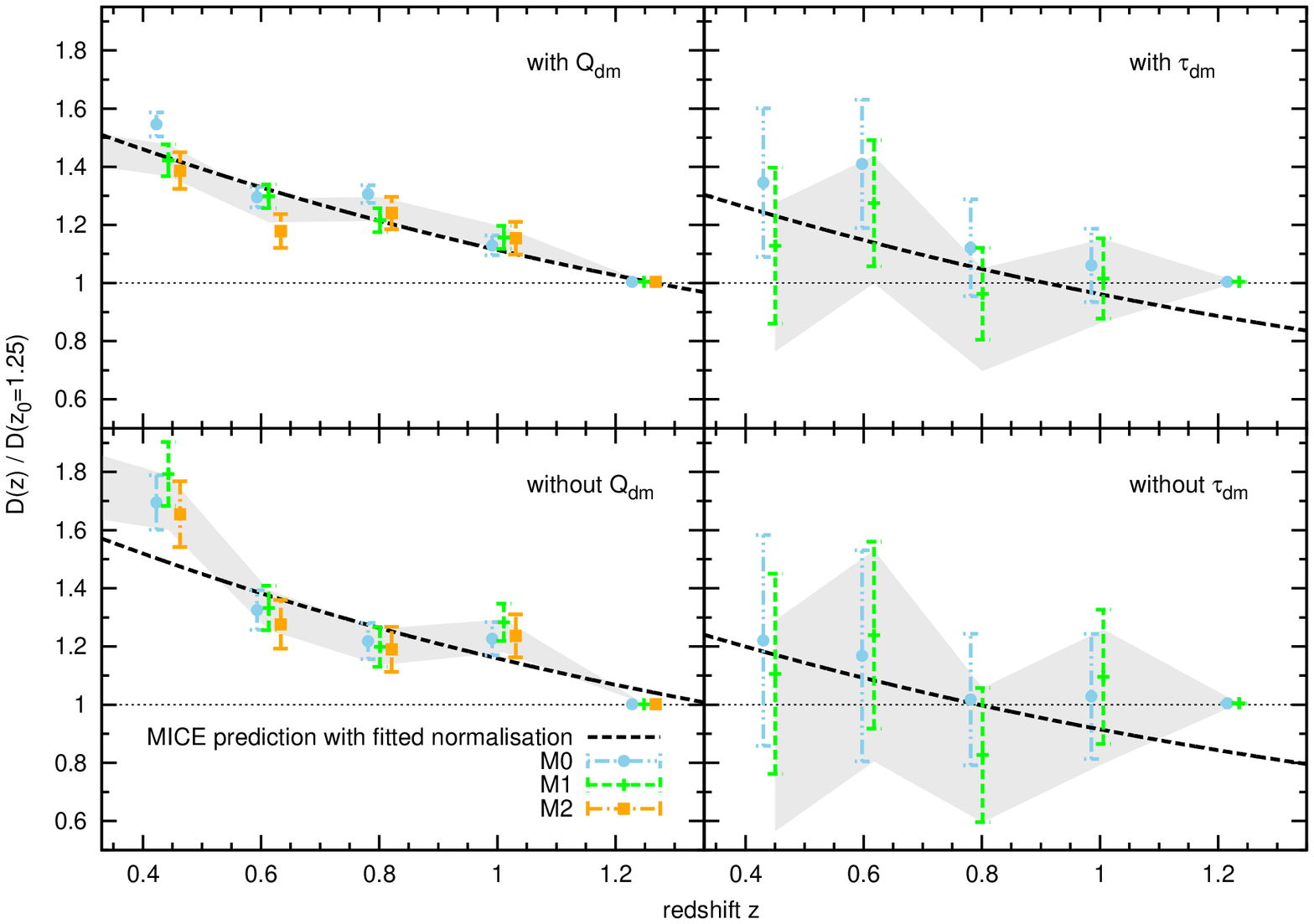}}
\caption{Growth factor measured from haloes in the mass samples M0, M1, M2 from the MICE-GC light cone. Measurements are normalised to be unity at the reference redshift $z_0=1.25$.
Symbols show results derived by using the same mass bin at redshift $z$ and $z_0$. Median results with median errors from combining all mass bins are shown as grey areas.
Measurements shown in the left  panels are derived using the linear bias from $Q(24,48,\alpha)$, while results in the right panel are based the linear bias from $\tau \equiv 3C_{12}-2S_3$.
The dashed line is the theoretical prediction from linear perturbation theory in equation (\ref{eq:growth_cosmology}) for the MICE-GC cosmology. Its normalisation was chosen to minimise
deviations from the median measurements and is therefore different in each panel. Results shown in the top panels are based on separate measurements of bias at each redshift, by comparing $Q$ or $\tau$ in haloes with the corresponding dark matter measurements in the same redshift bin (see \S3.4.1 and \S3.5.1). The results in the bottom panels are based on ratio measurements of bias at two redshift, by comparing $Q$ or $\tau$
at different redshifts (no dark matter is used, see \S3.4.1.2 and \S3.5.2).}
\label{fig:results8}
\end{figure*}

\subsubsection{Growth factor measured with $Q_{dm}$ and $\tau_{dm}$} 

\label{d_comparison1}

Fig. \ref{fig:results8} shows measurements of the growth factor $D$, derived from the mass samples M0, M1 and M2 in the MICE-GC light cone.
 Symbols denote results, which were derived by using the same mass bins at both redshifts, $z_0$ and $z$. Exploring the variation of our results for
different choices of mass bins, we measure the growth factor from all combinations of mass bins. The median growth factor and the median error from
all combinations are shown as grey shaded areas in the same figure.
The top panels show results, derived by using the bias parameters $b_Q$  and $b_\tau$ (left and right respectively), which were
measured at each redshift separately from equation (\ref{eq:b1c2_q3auto}) and (\ref{biaslin}).
This approach requires the knowledge or modelling of the dark matter $Q_{dm}$ and $\tau_{dm}$.
Note that we normalised all measurements with respect to the highest redshift bin by setting $z_0=1.25$ in equation (\ref{eq:D_from_2pc}).
This allows us to have a normalisation, which is performed as much as possible in the linear regime and with the lowest possible sampling variance.
The measurements are compared to the theoretical prediction from equation (\ref{eq:growth_cosmology}), shown as dashed lines in the same figure.
We also show that combining different halo populations (halo mass) at different redshifts we obtain results which are consistent with those derived by
following the same halo population across the considered redshift range. To be independent of the normalisation we $\chi^2$-fit the normalisation
of the predictions to the median measurements from all mass sample combinations (i.e halo populations). 

Our results in Fig. \ref{fig:results8} show that the growth factor, measured with the bias from the third-order statistics, decreases with redshift, as expected from
predictions for the linear growth factor. In the case of $Q$ (top left panel of Fig. \ref{fig:results8}) the good agreement between measured and predicted growth
factor is remarkable since the bias estimation, on which the measurement is based on, shows a $30\%$ over estimation (see left panel of Fig. \ref{fig:results6}).
We explain this finding by a cancellation in the bias ratio $\hat b_Q$ of the multiplicative factor by which $b_Q$ is shifted away from $b_{\xi}$.
This cancellation also happens for the median results from all mass bin combination, since this multiplicative factor is similar for all masses and redshifts.
Fluctuations of the growth factor measurements at high redshifts probably result from fluctuations in the bias measurements.
We expect an additional uncertainty in the growth measurement from the resolution of the density grid, used for computing $b_Q$
(see Appendix \ref{sec:app}). From equation (\ref{eq:D_from_2pc}) we estimate that the $5\%$ resolution error in $b_Q$ propagates into $D$ as an error of below $10\%$.

The growth factor measurements based on the $\tau$ estimator (top right panel) are significantly noisier than the $b_Q$ estimation, as expected from the linear bias estimations displayed in
Fig. \ref{fig:results6}. Note that the growth factor, measured via $b_{\tau}$, 
appears to be more strongly biased at higher mass ranges. However, this is only the propagation of the statistical fluctuation of the bias measurements into the higher redshift bin, which is used to
normalise the measured growth factor in all the other redshift bins (see Fig. \ref{fig:results6}). Thus, we can conclude the $\tau$ estimator provides an unbiased estimate of the growth factor.

\subsubsection{Growth factor measured without $Q_{dm}$ and $\tau_{dm}$} 
\label{d_comparison2}

In the bottom panels of Fig. \ref{fig:results8} we show the growth factor measurements based on the new approach, which uses the bias ratios $\hat{b}_Q$  and $\hat{b}_{\tau}$ derived from
equations (\ref{eq:bD_q3auto}) and (\ref{bgrow}) (left and right respectively). This means that we compare the statistical properties only of the halo density field at different redshifts, without
requiring knowledge about the dark matter quantities $Q_{dm}$, $\tau_{dm}$. As in the top panels, the symbols denote measurements using the same
mass bins at both redshifts, while median growth measurements and errors from all mass bin combinations are shown as grey shaded areas.

We find for both estimators slightly larger deviations from the linear theory compared to the results from the separate bias measurement, shown in the upper panels. 
This discrepancy tends to be larger as the redshift is decreasing, possibly due to three effects: 
i) noise in the measurements of third-order galaxy (or halo) correlations ($Q_g$, $\tau_g$) enters twice,
ii) non-linearities in the dark matter field become stronger at small redshift,
iii) sampling variance does not cancel out since the two halo correlations, on which the measurement is based on, come from different redshifts.

In practice that  last point iii) will also affect the first method, which uses  $Q_{dm}$ and $\tau_{dm}$
to get the absolute bias at each redshift. In our analysis, sampling variance cancels out between redshifts
because we use  $Q_{dm}$ and $\tau_{dm}$ measured in the same simulation where we measure the corresponding
halo values  $Q_{g}$ and $\tau_{g}$. In the analysis of a real survey this cancelation will not occur since one needs
to use models for $Q_{dm}$ and $\tau_{dm}$.

\begin{figure}
\centerline{\includegraphics[width=180mm, angle=270]{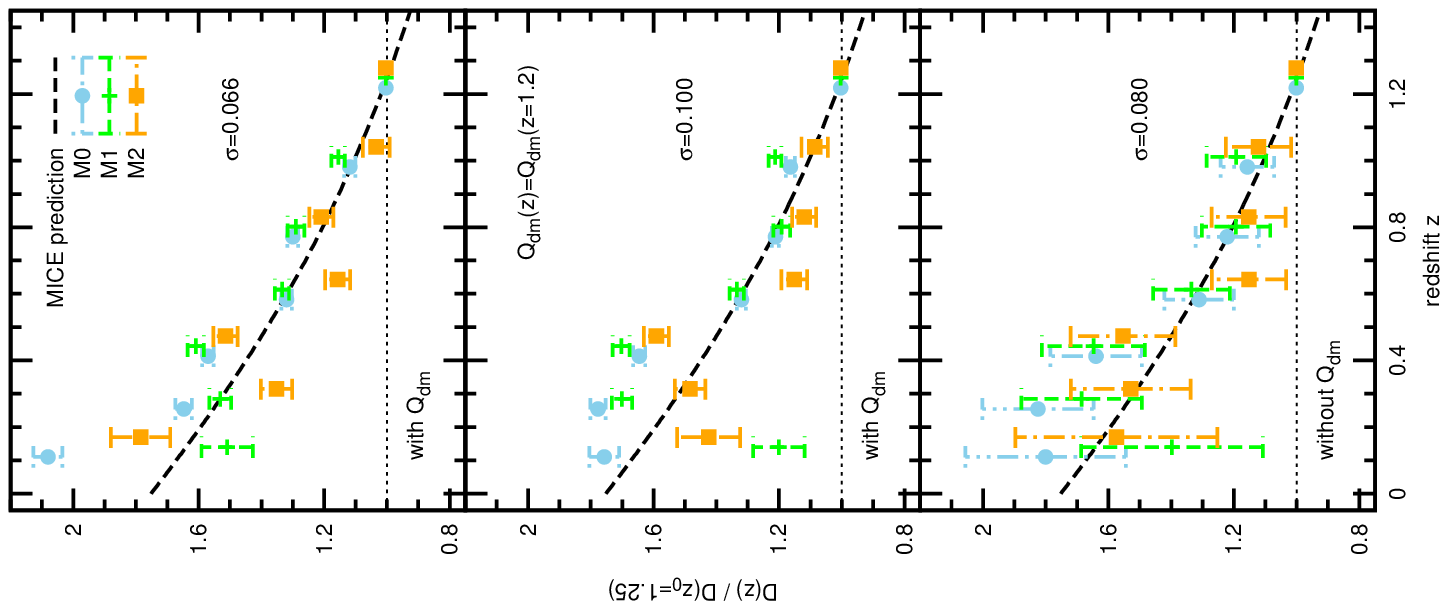}}
\caption{Impact of sampling variance on the linear growth factor, measured with $Q(24,48,0< \alpha <60)$ and normalised at $z_0=1.25$. The angular range
excludes widely opened triangles which are expected to be more strongly affected by sampling variance. The top panel shows $D(z)$ estimated from
separate measurements of bias at each redshift, by comparing $Q$ in haloes with the corresponding  dark matter measurements
in the same redshift bin. The central panel shows the same measurements when we use the dark matter measurements from $z_0$
for all redshifts instead. In this case sampling variance between halo and dark matter fluctuations does not canceled out, as it does in the top panel.
Results in the bottom panel are based on ratio measurements of bias at two redshift, by comparing $Q$ at different redshifts (no dark matter is used).
Quantifying the scatter we show the standard deviation $\sigma=\sqrt{\langle(D-D_{PT})^2\rangle}$, where $D_{PT}$ is the predicted growth factor
(shown as black dashed line) and $\langle  ... \rangle$ denotes the mean over all redshifts and mass samples.}
\label{fig:results9}
\end{figure}

We demonstrate this effect in Fig. \ref{fig:results9}. To study how the fitting range affects the growth estimate
we now restricted to opening angles between $0<\alpha<60$ degree, excluding large triangles as we discuss later. The top panel shows the growth,
measured with the separate bias estimates of $b_Q$, as shown on the top panel of Fig. \ref{fig:results8}. In the
central panel we show the more realistic growth measurements based on the same approach,
but instead of using the dark matter measurements in the same redshift as the halo measurements, we always use the dark matter
results of $Q$ from the highest redshifts bin $z_0=1.25$ (which is in good agreement with the results from the comoving output, as it contains more volume
than the other redshift bins). Here, the sampling variance does not cancel, since the dark matter and halo correlations are measured at different redshifts. This results in a larger
scatter in the central panel than in the top panel. Quantifying this scatter with respect to the predictions as $\sigma=\sqrt{\langle(D-D_{PT})^2\rangle}$
confirms the visual impression (values are shown in  Fig. \ref{fig:results9}). The latter approach in the central panel corresponds more closely to how the first
method would be applied in a real survey: i.e. assuming a cosmology to run the dark matter model and running a simulation for that cosmology (sampling
variance will not cancel as the simulation has different seeds than the real Universe). These latter growth measurements are distributed in a similar way
around the theoretical predictions as the results  derived from the ratio bias approach using $Q$ and $\xi$ at the redshift $z$ and redshift $z=1.25$, shown in the bottom panel
of Fig. \ref{fig:results9}.
This demonstrates that most of the difference between the top and bottom panels of Fig. \ref{fig:results9} comes from the artificial
sampling variance cancelation in the top panel. 

The restriction of the opening angles to $0<\alpha<60$ excludes widely opened triangles and is a
possibility to decrease the impact of sampling variance on the measurements. However, comparing the growth
in the bottom panel of Fig. \ref{fig:results9} to the corresponding measurements for $0<\alpha<180$
in the bottom left panel of Fig. \ref{fig:results8} we find no significant improvement of the growth measurements
from restriction of the opening angles possibly due to larger errors. The latter result from the smaller number
of triangles.

\subsubsection{Growth rate measured with and without $Q_{dm}$}
\label{f_comparison}

We derived the growth rate $f$ from the measured growth factor $D$ via equation (\ref{growratedis}).
The product $f \sigma_8$ can also be probed by redshift space distortions, while our measurements represent an additional, independent approach to the growth rate.
Especially at higher redshifts, where growth rate measurements via redshift space distortions are difficult to obtain, such additional information is valuable.
Besides the relation to redshift space distortions another advantage of the growth rate with respect to the growth factor is that it is independent of the normalisation.
The latter cancels out in the ratio of the growth factors, which appears in equation (\ref{growratedis}).

However, measuring $f$ via $D$ at a given redshift is not straightforward, since it depends on measurements at two
different redshift. We derived $f$ at the redshift bin $z_i$ from growth factor measurements at $z_{i+1}$ and $z_{i-1}$. This approach is motivated
by the fact that the redshift bins have equal width in comoving space. Constrains of cosmological parameters with such measurements would
require a more careful treatment of the assigned redshift. The employed growth factors are the median results, derived via $Q$ from all mass
combinations, which are shown as grey areas in the left panels of Fig. \ref{fig:results8}.

\begin{figure}
\centerline{\includegraphics[width=80mm,angle=270]{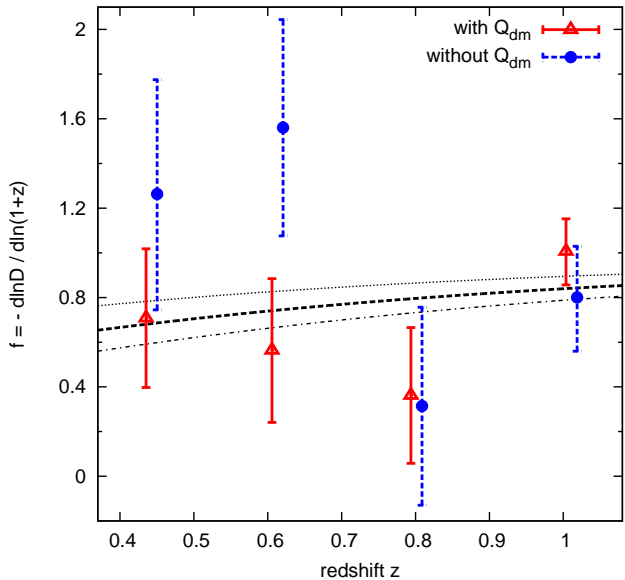}}
\caption{ Growth rate $f$, estimated from the median measurements of $D$ from all mass bin combinations
(shown as grey areas in the left panels of Fig. \ref{fig:results8}) via equation (\ref{growratedis}).
MICE predictions, derived from equation (\ref{growing2}) with $\Omega_m=0.25$ and $\gamma=0.55$
are shown as thick dashed lines. By changing the values of $\gamma$ to $0.35$ and $0.75$, we derive the
predictions for different cosmologies, shown as dotted and dash-dotted lines, respectively.
The large errors could be decreased by measuring the bias $b_Q$ using a combination of different triangle configurations.}
\label{fig:results10}
\end{figure}

The results for $f$ from $D$, measured with and without $Q_{dm}$, are shown in Fig. \ref{fig:results10}.
In both cases the measurements are strongly scattered around the theoretical predictions for the MICE simulation,
while the scatter is stronger for results derived without $Q_{dm}$. The increased scatter at lower redshifts probably
results from the smaller volume of the light cone, which causes stronger fluctuations in $D$ (see Fig. \ref{fig:results8}).
We also expect an uncertainty from the resolution of the density grid, used for computing $b_Q$ (see Appendix \ref{sec:app}).
From the equations (\ref{eq:D_from_2pc}) and (\ref{growratedis}) we estimate that the $5\%$ resolution error in $b_Q$ propagates
into $f$ as an error of below $10\%$.
Note that the errors that we find at high redshifts are comparable, or slightly better, than current errors from
redshift space distortions (RSD) in the anisotropic 2-pt correlation function. 
Note that measurements from RSD directly
constrain $f \times D$ and not $f$. Nevertheless, under some assumptions we can
also infer $f$ from RSD and the typical errors found
on SDSS, BOSS and WiggleZ are around 15-20\% \citep{C&G09,Blake2011,Tojeiro2012}, which are comparable
to the ones we find here.

The MICE prediction is computed from equation (\ref{growing2}) with $\Omega_m=0.25$
and $\gamma=0.55$. To compare the scatter in our measurements with variations of the growth rate for different cosmologies we also show predictions for
$\gamma=0.35$ and $0.75$. We find that our errors in the measurements are larger than the expected variations in the growth rate due to cosmology.
It would be worthwhile to conduct a similar comparison using larger mass bins and combining measurements from different scales and configurations of $Q$,
to decrease the error, but the goal here is just to demonstrate the
possibility of such measurements and the advantage of using it.
  
\begin{figure}
\centerline{\includegraphics[width=80mm]{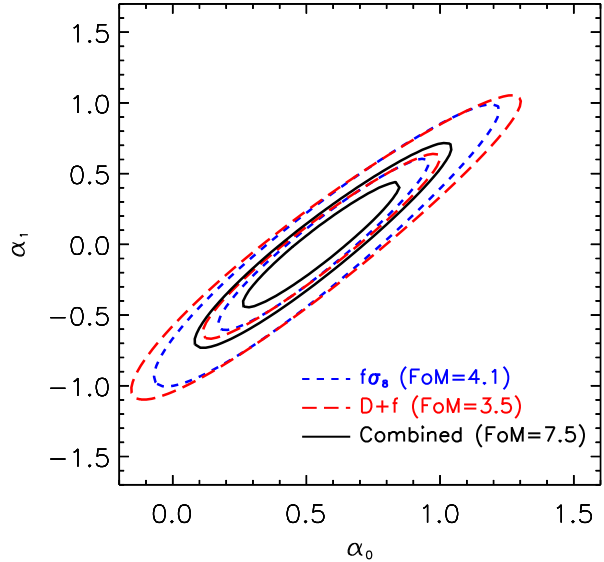}}
\caption{ Comparison between estimating the growth-index ($\alpha_0$,  $\alpha_1$) from the measurement of the redshift-space distortion parameter $f\sigma_8$ (blue short dashed line) with
the independent measurement of $f$ and $D$ (red long dashed line). For both we simulated $7$ measurements corresponding to redshifts $0.3$, $0.5$, $0.7$, $0.9$, $1.1$, $1.3$ and $1.5$ 
with a $12$\% relative error at each redshift. The Figure-of-Merit (FoM) is computed by taking the inverse of the area enclosed into the $1$-$\sigma$ contour (inner contour of each ellipse). The solid line shows the combination of the two. }
\label{fig:results11}
\end{figure}

We also studied a test case where we compare the
inference power of a measurement of $f\sigma_8$ with a measurement of
$f$ and $D$ independently. To do so we simulated seven measurements of
$f\sigma_8$ and of $f$ and $D$ at seven different redshifts (see
caption of Fig. \ref{fig:results11}) and we compared the constraining
power of the two methods in determining the growth index
$\alpha=\alpha_0 + \alpha_1\ln\Omega_m$ characterized by the two free
parameters $\alpha_0$ and $\alpha_1$.
We stress that this analysis is
purely illustrative, in the sense that we just compare the intrinsic
potential of each method, i.e. we assume the same number of
measurements at the same redshift and 
with the same relative errors. Fig. \ref{fig:results11} shows
that the constraining power is compatible for the two methods leading
therefore to an improvement close to a factor of $2$ when combining
them (black lines). However, note that in case of $f\sigma_8$ it is necessary to assume an external prior on the scalar amplitude of the initial power spectrum. We indeed should have taken a Gaussian prior on it \citep[see second column of Table 2 in][]{Ade13}, however assuming that any dark-energy or modifications of gravity would kick in between redshift $0$ and $10$, we equivalently assumed a Gaussian prior (with the same relative error) on the normalisation of the power spectrum $\sigma_8$ at redshift $10$. While for the method we propose, no priors on the scalar amplitude are assumed.


%%%%%%%%%%%%%%%%%%%%%%%%%%%%%%%%%%%%%%%%%%%%%%%%%%%%%%%%%%%%%
%%%%%%%%%%%%%%%%%%%%%%%%%%%%%%%%%%%%%%%%%%%%%%%%%%%%%%%%%%%%%
%%                                              DISCUSSION (Section V)                                             %%
%%%%%%%%%%%%%%%%%%%%%%%%%%%%%%%%%%%%%%%%%%%%%%%%%%%%%%%%%%%%%
%%%%%%%%%%%%%%%%%%%%%%%%%%%%%%%%%%%%%%%%%%%%%%%%%%%%%%%%%%%%%

\section{Summary \& Discussion}\label{sec:disc}

The amplitude of the transverse (or projected) two-point correlation of matter density fluctuations allows us to measure the growth factor $D$,
which can be used as a verification tool for cosmological models. Galaxies (in our study represented by haloes) are biased tracers of the full
matter field as their two-point correlation at large scales is shifted by a constant bias factor $b$ with respect to the matter two-point correlation.
This bias factor is fully degenerate with $D$. The reduced matter and galaxy third-order statistics are independent of $D$, while the galaxy
versions are sensitive to $b$. Combining second- and third-order statistics could therefore enable us to break the growth-bias degeneracy,
if the difference between the effective linear bias $b_1$ probed by both statistics is smaller than the errors required for the growth measurements.

In this paper we have tested these  assumptions and verified how well we can recover the true growth of the new MICE-GC
$\Lambda$CDM simulation \citep{mice1, mice2, mice3} with them.
We also further validate the MICE-GC simulation by comparing the linear
growth with the  two-point  matter correlation (Fig.2 and 3) and the different
third-order statistics of the matter field
to non-linear perturbation theory predictions (Fig.5 and 8).
In particular, previous analysis \citep{GFC} found a
mismatch between simulations and predictions for $C_{12}$
\citep{bernardeau96}, which we find here to originate from neglecting one
of the smoothing terms (i.e. $\beta_R$ in Eq.28). After taking
this into account, the MICE-GC simulations agrees well
with predictions for all redshifts (see Fig.8). This, therefore,
provides a validation of the approach adopted by \citet{bm} to
measure the linear galaxy bias using only galaxy clustering.

The main goal of this paper is to compare bias (and the resulting growth) measurements
from two different third-order statisticics proposed in the literature. 
One uses the reduced three-point
correlation $Q$ while the other uses a combination
of the skewness $S_3$ and two-point third-order correlator, $C_{12}$,
which is called $\tau\equiv 3C_{12}-2S_3$. 
We estimated these quantities from density fields of matter in the
MICE-GC simulation and those of haloes in different mass samples, 
expanding previous studies  significantly to  a wider range of masses
(between $5.8\times 10^{12}$ and $5\times 10^{14}$ $h^{-1}$M$_\odot$)
and redshifts  (between $0$ and $1.2$) with values of the linear bias $b_1$
between $0.9$ and $4$.

\begin{figure}
  \centering
   \includegraphics[width=85mm]{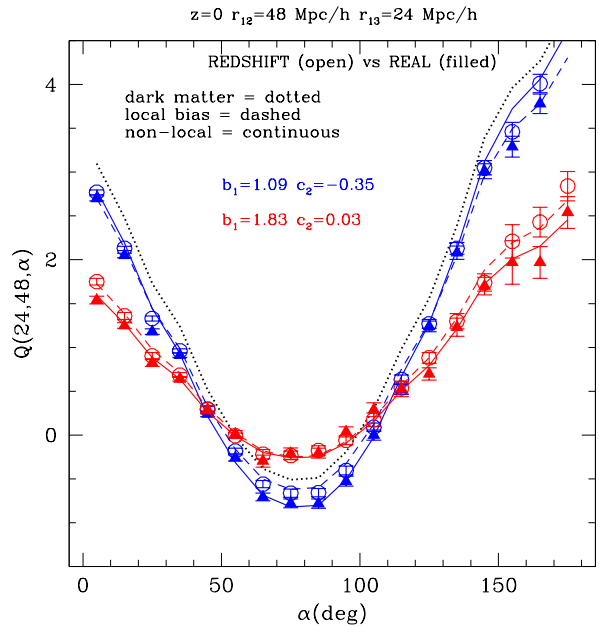}
   \caption{$Q$ for dark matter (dotted) and for halo samples
     (symbols) with two
     different mass thresholds:  $b_1 =b_\xi\simeq 1.09$ (blue)
     and $b_1 =b_\xi \simeq 1.83$ (red).  We compare results in real space (filled triangles) and redshift space
(open circles), which agree within the errors on these
large scales ($r_{12} = r_{13}/2 = 24$  $h^{-1}$Mpc at z=0).  Predictions are shown for both:
the local bias model (dashed lines) and non-local bias model (continuous). In both cases we have fixed $b_1=b_\xi$ and fit 
for $c_2$. }
 \label{fig:q3pz}
\end{figure}

Our results in Fig. \ref{fig:results4} show that the linear bias from $Q$, $b_Q$, systematically over estimates the linear bias from the two-point correlation, $b_{\xi}$,
by roughly $20-30\%$ at all mass and redshift ranges, whereas the linear bias from $\tau$, $b_\tau$, seems  to be an unbiased estimator at the price of decreased 
precision. Non-local contributions to galaxy bias, like tidal effects, are anisotropic and therefore could be more
important for $b_Q$  than for $b_\tau$, as $\tau$ is isotropic (i.e. it comes from higher-order
one- and two-point correlations, while $Q$ comes from three points). In Fig. \ref{fig:q3pz} we illustrate the different impacts of
the local bias model and the non-local model of \cite{chan12} with $\gamma_2=2(b_1-1.43)/7$  on $Q$ (dashed and solid lines respectively).
The non-local model seems to approximate $Q$ measurements from halo samples better but there are still some
discrepancies that we will explore in a separate analysis.
Besides non-local contributions to the bias model, further reasons for the difference between $b_Q$ and $b_{\tau}$ might be
that non-linear terms in the bias function and the matter field have different impacts on $Q$ and $\tau$.
In addition we found that estimations of the quadratic bias parameter
$c_2$ from $Q$ and $\tau$ can also differ significantly from each
other.

Understanding the differences between $b_{\xi}$, $b_Q$ and $b_{\tau}$ is crucial for constraining cosmological models
with observed third-order halo statistics. We will therefore deepen our analysis in a second paper by studying bias from
halo-matter-matter statistics, direct analysis of the halo versus matter fluctuations and predictions from the peak-background
split model to disentangle between non-linear and non-local effects on the different estimators.

For measuring the growth factor $D$ we have introduced a new method. This new method uses
the bias ratio $\hat{b}(z)=b(z)/b(z_0)$, derived directly from halo density fluctuations with
reduced third-order statistics. Its main advantage with respect to the 
approach of measuring $b(z)$ and $b(z_0)$ separately is that
it does not require the modelling of (third-order) dark matter
statistics. Instead, it works with the hypothesis that
\begin{enumerate}
\item the reduced dark matter three-point statistics is independent of redshift $z$
\item the bias ratio $\hat{b}(z)=b(z)/b(z_0)$ from two- and  three-point statistics is equal.
\end{enumerate}
The first assumption was tested in this study numerically, while the validity of the second follows directly
from our bias comparison.

In general the comparison between $D$ from perturbation theory with measurements from our new method and the standard approach reveals
a good agreement. In the case of $Q$  we explain this result by a cancellation of the multiplicative factor by which $b_Q$ is shifted away from $b_{\xi}$
in the bias ratio $\hat b_Q$. The growth factor measured with $\tau$ has larger errors than the results from $Q$ as a consequence of the larger errors
in the bias estimation. 

Our analysis shows that the new way to measure the growth factor from bias ratios is competitive with the method based on two separate bias measurements.
While having larger errors the new method has the advantage of requiring much weaker assumptions on dark matter correlations than the standard method
and therefore provides an almost model independent way to probe the growth factor of dark matter fluctuations in the Universe.

We demonstrated that besides the growth factor, $D$,   the growth rate of matter, $f$, can also 
be directly measured from the galaxy (or halo) density fields
with bias ratios from third-order statistics. This provides an alternative method to derive the growth rate,
which is usually obtained from velocity distortions probed by the anisotropy of the two-point correlation function (RSD). 
The typical errors found on SDSS, BOSS and WiggleZ using RSD are around 15-20\% \citep{C&G09,Blake2011,Tojeiro2012}, which are comparable
to the ones we find in Fig. \ref{fig:results10} when considering the high redshift bins (20\%).

Given that the two methods explored here use different information
from higher-orders correlation ($Q$ uses the shape, while $\tau$ uses
collapse configurations) one can reasonably guess that the two methods are not
strongly correlated. So a possible strategy would be to use the $Q$ method (more
precise) to measure the (velocity) growth rate and, in parallel, to use
the $\tau$ method to extract the growth factor. This would help to
break degeneracies between cosmological parameters in different
gravitational frameworks. 

Our analysis is performed in real space to have clean conditions for comparing different bias and growth estimates.
This is a good approximation for the reduced higher-order correlations on the large scales considered in this study,
as measurements in redshifts space  always seem to be within one sigma error of the corresponding real space result 
(see Fig. \ref{fig:q3pz}).
Note how the small, but systematic, distortions in redshifts space seem to agree even better with the local bias model
than in real space on the largest scales.

Applying the methods described above to obtain accurate bias and growth measurements from observations
will require additional treatment of redshifts space distortions or projection effects.
Two possible paths could be followed. In a three dimensional analysis redshifts space distortions need to be modeled
\citep[e.g.][]{gaztascoc}. The projected three-point  correlation can
also be studied separated by in redshift bins \citep{friga1999,Buchalter,zheng}.
Both ways will result in larger errors, but we do not expect this to be a limitation
because our error budget is totally dominated by the uncertainty in the bias.
A more detailed study of this issue is beyond the scope of this paper and will be presented elsewhere.
Mock observations,  like the galaxy MICE catalogues \citep[see][]{mice2,Carretero2014}
should be used to test the validity of such growth measurements under more realistic conditions.

%%%%%%%%%%%%%%%%%%%%%%%%%%%%%%%%%%%%%%%%%%%%%%%%%%%%%%%%%%%%%
%%%%%%%%%%%%%%%%%%%%%%%%%%%%%%%%%%%%%%%%%%%%%%%%%%%%%%%%%%%%%
%%                                                ACKNOWLEDGEMENT                                               %%
%%%%%%%%%%%%%%%%%%%%%%%%%%%%%%%%%%%%%%%%%%%%%%%%%%%%%%%%%%%%%
%%%%%%%%%%%%%%%%%%%%%%%%%%%%%%%%%%%%%%%%%%%%%%%%%%%%%%%%%%%%%
\section*{Acknowledgements}
 
Funding for this project was partially provided by the Spanish Ministerio de Ciencia e Innovacion (MICINN), project AYA2009-13936,
Consolider-Ingenio CSD2007- 00060, European Commission Marie Curie Initial Training Network CosmoComp (PITN-GA-2009-238356) and research
project 2009- SGR-1398  from Generalitat de Catalunya.
JB acknowledges useful discussions with Christian Marinoni and support of the European Research
Council through the Darklight ERC Advanced Research Grant (\#291521).
KH is supported by beca FI from Generalitat de Catalunya. He also acknowledges the Centro de Ciencias de Benasque Pedro Pascual
where parts of the analysis were done.
The MICE simulations have been developed at the MareNostrum supercomputer (BSC-CNS) thanks to grants AECT-2006-2-0011
through AECT-2010-1-0007. Data products have been stored at the Port d'Informaci— Cient'fica (PIC).

%%%%%%%%%%%%%%%%%%%%%%%%%%%%%%%%%%%%%%%%%%%%%%%%%%%%%%%%%%%%%
%%%%%%%%%%%%%%%%%%%%%%%%%%%%%%%%%%%%%%%%%%%%%%%%%%%%%%%%%%%%%
%%                                                        REFERENCES                                                     %%
%%%%%%%%%%%%%%%%%%%%%%%%%%%%%%%%%%%%%%%%%%%%%%%%%%%%%%%%%%%%%
%%%%%%%%%%%%%%%%%%%%%%%%%%%%%%%%%%%%%%%%%%%%%%%%%%%%%%%%%%%%%

\bibliographystyle{mnbst}
\bibliography{ref_hbgcfc_rev3.bib}

%%%%%%%%%%%%%%%%%%%%%%%%%%%%%%%%%%%%%%%%%%%%%%%%%%%%%%%%%%%%%
%%%%%%%%%%%%%%%%%%%%%%%%%%%%%%%%%%%%%%%%%%%%%%%%%%%%%%%%%%%%%
%%                                                        APPENDIX                                                    %%
%%%%%%%%%%%%%%%%%%%%%%%%%%%%%%%%%%%%%%%%%%%%%%%%%%%%%%%%%%%%%
%%%%%%%%%%%%%%%%%%%%%%%%%%%%%%%%%%%%%%%%%%%%%%%%%%%%%%%%%%%%%

\appendix 
\section{impact of covariance and resolution in $Q$ on linear bias estimation}\label{sec:app}

\begin{figure}
  \centering
   \includegraphics[width=225mm, angle = 270]{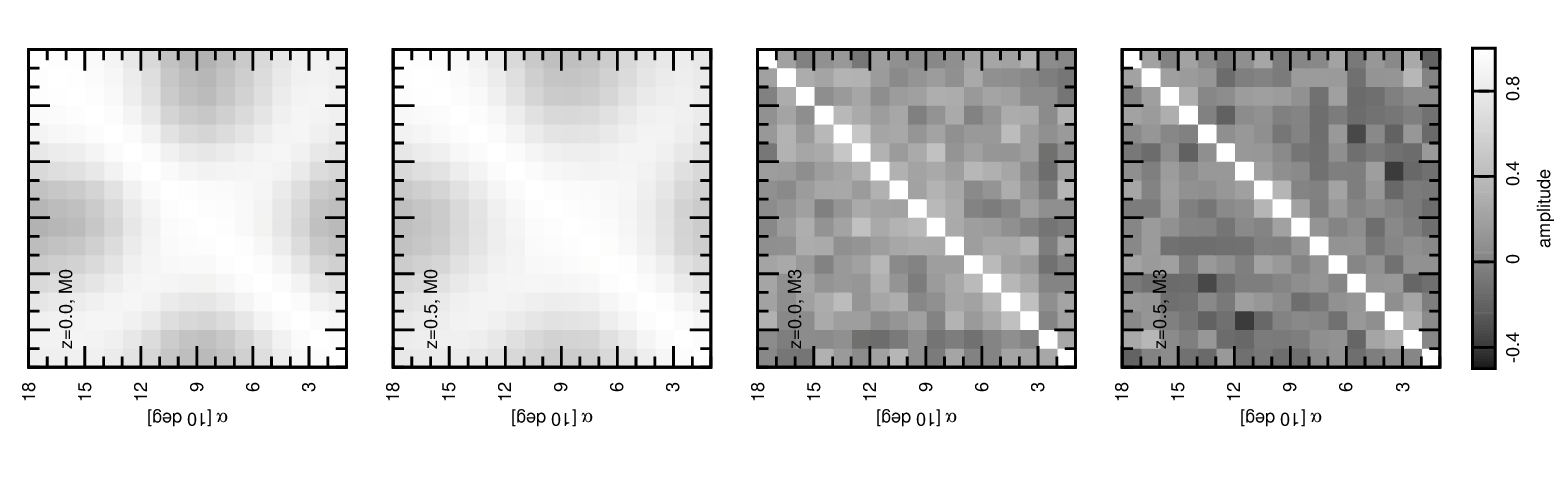}
   \caption{Normalised covariance $C_{ij}$ between the $18$ opening angles of $Q(24,48,\alpha$) for the mass samples M0 and M3 at redshifts $z=0.0$ and $z=0.5$.}
 \label{fig:covar}
\end{figure}

The jackknife estimation of the covariance matrix for $Q$, $C_{ij}$, measured for different opening angles $\alpha_i$, is a potential
reason for the discrepancy between the linear bias from two- and reduced three-point correlations ($b_{\xi}$ and $b_Q$ respectively).
Studying how strong our bias estimation is affected by the covariance matrix we compare $b_Q$ derived with the jackknife
covariance matrix to results measured without taking covariance into account, i.e. by setting  $C_{ij} = \delta_{ij}$.

We show the covariance matrixes of $Q$ with ($24$,$48$) configurations (see Table \ref{triangle_configs} for details)
in Fig. \ref{fig:covar}. For the low mass sample M0 $C_{ij}$ has a similar shape as results of \citet{gaztascoc}.
The off-diagonal elements are close to unity, which corresponds to $Q$ at intermediate opening angles
($70-80$ deg) having covariance with values at large and small angles. For the high mass sample M3 the covariance
is dominated by noise.

\begin{figure*}
  \centering
   \includegraphics[width=150mm, angle = 270]{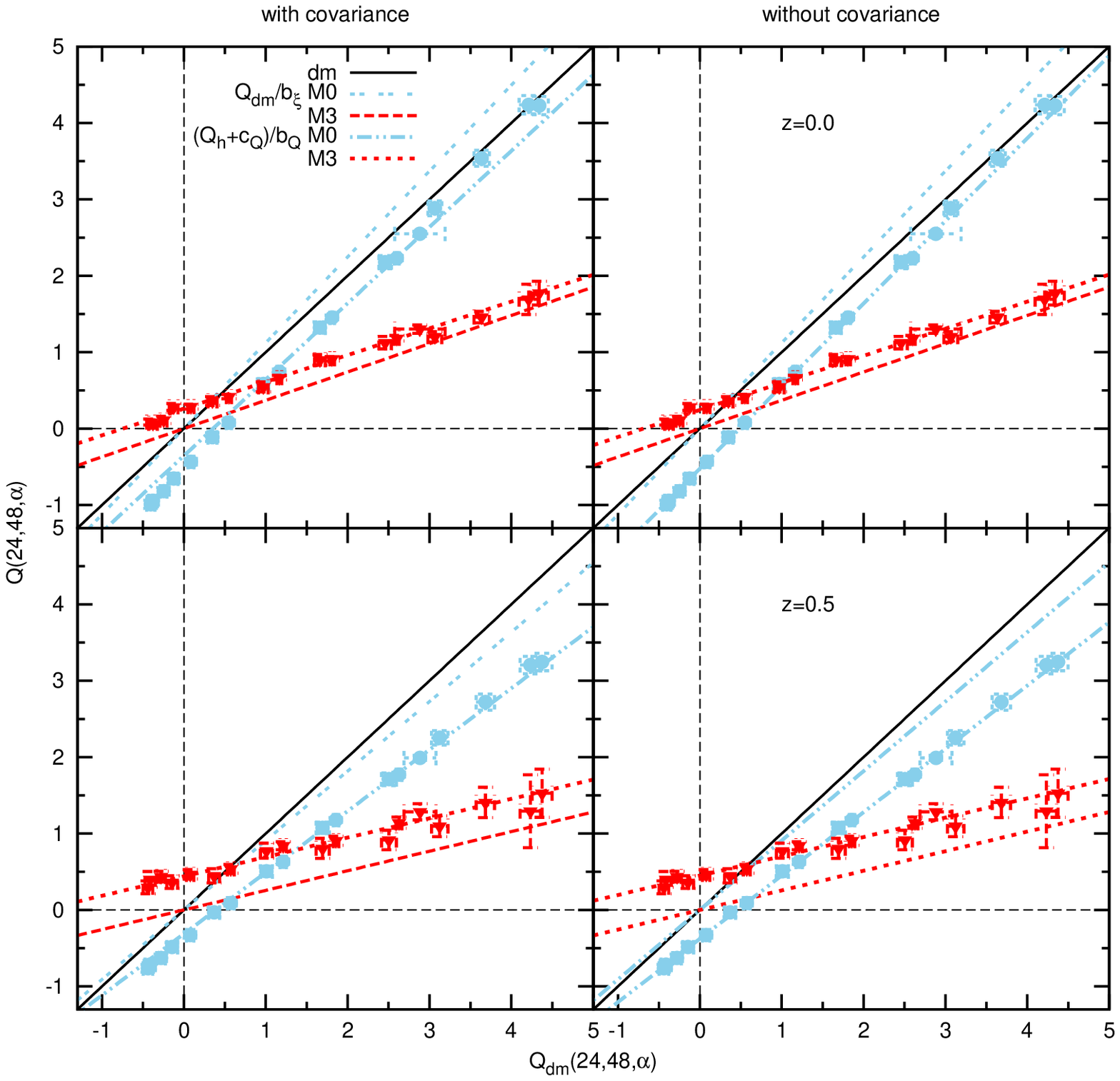}
   \caption{$Q$ for the high and low galaxy (or halo) mass samples M0 and M3 versus $Q$ for dark matter at the corresponding opening angle.
   Dotted and dash-dotted lines are $\chi^2$-fits to the $Q_g$-$Q_{dm}$ relation expected from perturbation theory (equation (\ref{eq:b1c2_q3auto})).
   The fits were performed with and without taking the jackknife covariance of $Q_g$ between different opening angles into account (left and right panel respectively).
  Long-dashed and double dotted lines show expected results for a linear bias model, using the linear bias measurement from the two-point correlation, $b_{\xi}$.
  Bottom and top panels show results at redshift $z=0.0$ and $z=0.5$.}
 \label{fig:qvsqdm_covnocov}
\end{figure*}

Examples for how well the fits to equation (\ref{eq:b1c2_q3auto}) match the measured relation between
$Q_g$ and $Q_m$ are shown in Fig.Ê\ref{fig:qvsqdm_covnocov}. The fits are shown as coloured line,
while their inverse slope corresponds to $b_Q$ and the crossing point with the y-axis marks $c_Q/b_Q$.
Especially for the low mass sample at redshift $z=0.0$ bias measurements performed without jackknife
covariance seem to deliver better fits to the measurements.
According to the covariance of Fig. \ref{fig:covar} the fit allows deviations in the intermediate angle that are compensated with
correlated deviations at large and small scales. This produces a change in the value of the fitted bias.
Whether this change is correct or not depends on whether the covariance is correct or not.
For the higher mass samples and for both mass samples at redshift $z=0.5$ results derived with and without
covariance appear to be more similar. In these cases the off-diagonal regions of the covariance matrixes are
less pronounced, especially for the high mass sample.

In the same figure we compare these fits to results, expected for a linear bias model with $b_Q=b_{\xi}$ and $c_Q$=0.
For the low mass sample M0 at $z=0.5$ we find that the slopes from such a model match neither the measured $Q_g-Q_{dm}$
relations nor the fits to these measurements from equation (\ref{eq:b1c2_q3auto}). In all other cases differences between the
slopes expected from the linear bias model and the measured $Q_g-Q_{dm}$ relations are less obvious.

\begin{figure*}
  \centering
   \includegraphics[width=145mm, angle = 270]{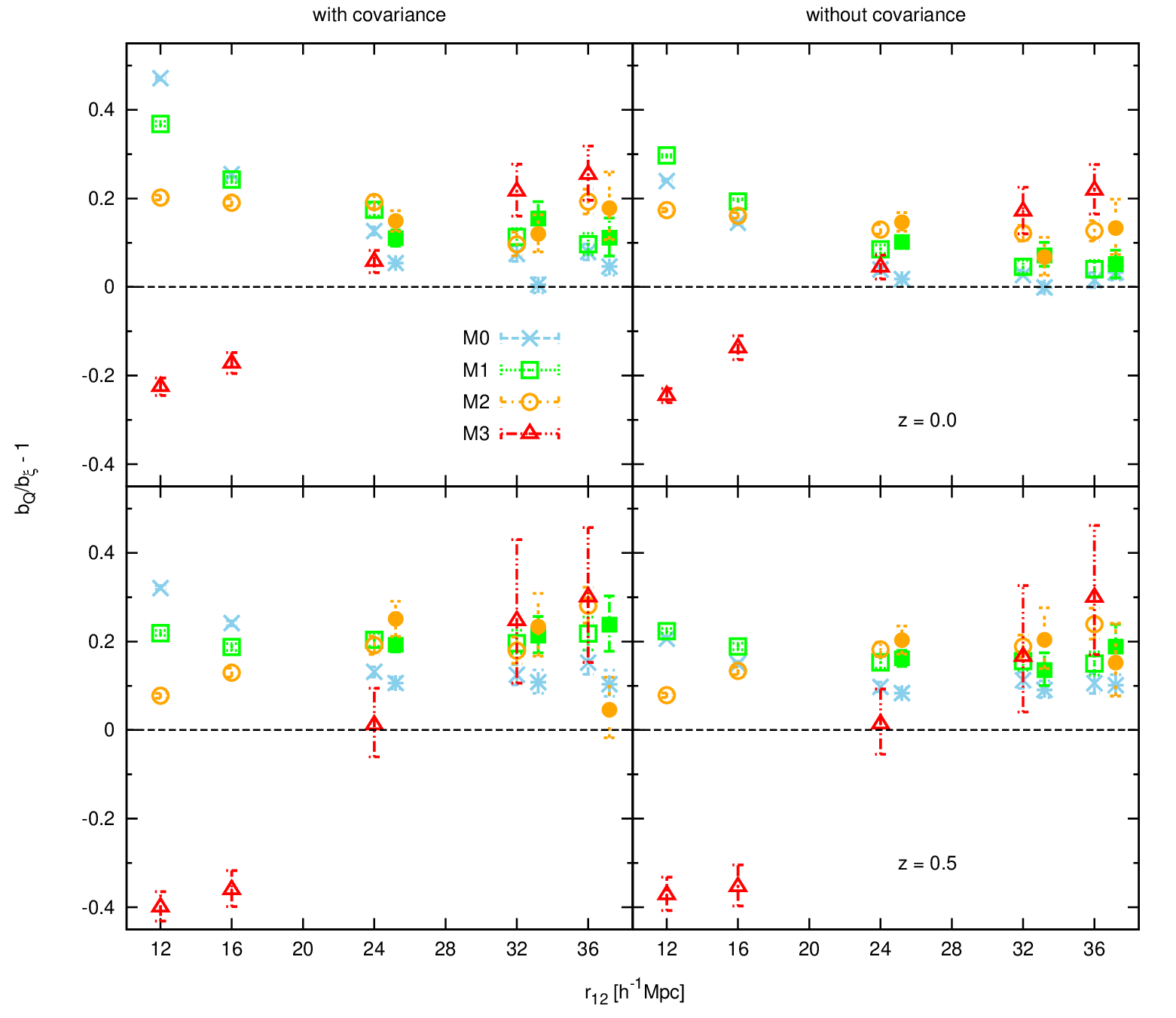}
   \caption{Relative deviations between the linear bias parameters $b_{\xi}$ and $b_Q$ derived from two-point and reduced three-point correlations respectively.
   $b_{Q}$ was derived using triangles with $r_{13}/r_{12}=2$ configurations, while the $r_{12}$ values are shown on the x-axis. Left and right panels show, respectively,
   results obtained with and without taking the jackknife covariance between $Q$ at different opening angles into account. Bottom and top panels show results at redshift $z=0.0$ and $z=0.5$
   respectively.
   Open symbols show results from $Q$ using triangles consisting of $3$ and $6$ grid cells per leg,
  while the triangle scale is increased by increasing the grid cell size. Closed symbols (slightly shifted to larger scales for clarity) show results from using the smallest available
  grid cell size of $4$ $h^{-1}$Mpc, while the triangle scale is increased by increasing the number of grid cells per leg. In the latter case also the shells use to contract triangles
  are chosen to be thinner (see Table \ref{triangle_configs} for details). Results for M3 then become very noisy and are therefore not shown.}
 \label{fig:bq_covnocov}
\end{figure*}

A comparison between $b_{\xi}$ and $b_Q$ measured with and without covariance at different scales is given in Fig. \ref{fig:bq_covnocov}.
Bias measurements from $Q$ performed without covariance tend to lie closer to the linear bias from the two-point correlation
$\xi$, while the overall trend towards overestimation remains. The fact that $b_Q$ measurements at large scales
for low mass samples at $z=0.0$, measured without covariance, lie very close to the corresponding $b_{\xi}$ values suggests that,
besides the jackknife estimation of the covariance, departures from the quadratic bias model for strongly biased halo samples
with high mass at high redshift contribute in a non neglectable way to the $b_{\xi}$ and $b_Q$ discrepancy. 
Furthermore non-local contributions to the bias model are expected to be strongest for such highly bias samples \citep{chan12}.
We concluded that the discrepancy between $b_Q$ and $b_\xi$ cannot be only due to uncertainties in the covariance matrix estimation.

In Fig. \ref{fig:bq_covnocov} we also show examples for $b_Q$ derived using smaller grid cell sizes and thinner shells to construct the triangles.
These computations are more expensive than those based on larger grids cells, but closer to the theoretical picture. We find that the bias values
change in most cases by around $5$ percent. These changes can be driven by changes in the amplitude of $Q$, but also by changes of the
covariance matrix. Especially for high mass samples and at large scales and higher redshift the amplitude of $Q$ becomes more noisy, which
can result in larger $\chi^2$ values in the fit. The covariance becomes more diagonal since for smaller grid cells the different triangle
opening angles are more independent of each other.

\begin{table*}
\centering
  \caption{Characteristics of the triangles used to measure $Q$. $r_{12}$ and $r_{13}$ are the fixed sizes of two triangle legs. $n_{12}$ and $n_{13}$ are
  the numbers of cubical grid cells per triangle leg. $dn$ is the tolerance for triangle leg sizes in units of grid cells used to define shells for constructing 
 triangles. $l_{cell}$ is the size of the cubical grid cells. The Figures showing results based on the different characteristics are given in the right column.}
  \begin{tabular}{c  c c c c c c l}
	$r_{12}$	&	$r_{13}$&	$l_{cell}$	&	$n_{12}$	&	$n_{13}$	&	$dn_{12}$&	$dn_{13}$	&	Fig. \\ 
$h^{-1}$Mpc & $h^{-1}$Mpc& $h^{-1}$Mpc & &  & & \\
	\hline
	$12$		&	$24$	&	$4$	&	$3$	&	$6$	&	$0.5$	&	$0.5$	&	\ref{fig:3pc1}, \ref{fig:3pc2}, \ref{fig:3pc3}, \ref{fig:bq_covnocov} \\
	$16$		&	$32$	&	$4$	&	$4$	&	$8$	&	$0.5$	&	$0.5$	&	\ref{fig:3pc3}, \ref{fig:bq_covnocov}\\
	$24$		&	$48$	&	$8$	&	$3$	&	$6$	&	$0.5$	&	$0.5$	&	\ref{fig:3pc2}, \ref{fig:3pc3}, \ref{fig:results4}, 
	\ref{fig:results6}, \ref{fig:q3pz}, \ref{fig:covar}, \ref{fig:qvsqdm_covnocov}, \ref{fig:bq_covnocov}\\
	$24$		&	$48$&	$4$	&	$6$	&	$12$	&	$0.1$	&	$0.1$	&	\ref{fig:3pc1}, \ref{fig:bq_covnocov}\\
	$32$		&	$64$	&	$8$	&	$4$	&	$8$	&	$0.5$	&	$0.5$	&	\ref{fig:3pc3}, \ref{fig:bq_covnocov}\\
	$32$		&	$64$&	$4$	&	$8$	&	$16$	&	$0.1$	&	$0.08$	&	\ref{fig:bq_covnocov}\\
	$36$		&	$72$	&	$12$	&	$3$	&	$6$	&	$0.5$	&	$0.5$	&	\ref{fig:3pc3}, \ref{fig:bq_covnocov}\\
	$36$		&	$72$	&	$4$	&	$12$	&	$18$	&	$0.05$	&	$0.06$	&	\ref{fig:bq_covnocov}\\
   \end{tabular}
    \label{triangle_configs}
\end{table*}

\end{document}